\documentclass[12pt]{article}
\usepackage{jheppub}

\usepackage{amsmath,amssymb,amsbsy}
\usepackage{graphicx,color}
\usepackage{placeins}
\usepackage[dvipsnames]{xcolor}
\usepackage{cancel,slashed,mathrsfs}
\usepackage{orcidlink}
\usepackage{hyperref}
\usepackage{framed}
\usepackage{cleveref}
\usepackage{bm}
\DeclareMathOperator{\diag}{diag}

\definecolor{shadecolor}{rgb}{0.90,0.90,0.90}

\graphicspath{{./plots/}}
\renewcommand{\d}{\mathrm d}
\newcommand{\Disc}{\operatorname{Disc}}

\newcommand{\EEC}{\operatorname{EEC}}
\newcommand{\eec}{\operatorname{eec}}
\newcommand{\bz}{\bar z}
\newcommand{\bw}{\bar w}
\newcommand{\be}{\begin{eqnarray}}
\newcommand{\ee}{\end{eqnarray}}
\newcommand{\bea}{\begin{eqnarray}}
\newcommand{\eea}{\end{eqnarray}}

\newcommand{\Li}{\mathrm{Li}}

\title{Energy-energy correlators in the large-charge limit of $\mathcal{N}=4$ super Yang–Mills theory}
\author[1,2]{Alessandro Georgoudis\,\orcidlink{0000-0002-1601-4218},}
\author[1]{Max Jackson\,\orcidlink{0009-0000-7915-4704},}
\author[1]{and Congkao Wen\,\orcidlink{0000-0002-5174-1576}}
\affiliation[1]{Centre for Theoretical Physics and Astronomy, Department of Physics and Astronomy,\\
  Queen Mary University of London, Mile End Road, London E1~4NS, United Kingdom}
\affiliation[2]{Department of Mathematics, City St George's, University of London,\\
  Northampton Square, London EC1V~0HB, United Kingdom}
\preprint{QMUL-PH-26-38}

\abstract{%
We study the energy-energy correlator (EEC) in heavy states of four-dimensional $\mathcal N=4$ super Yang--Mills theory in the large-charge 't Hooft limit, in which the dimension $\Delta$ of the heavy operator (or, equivalently, its $R$-charge) is taken to infinity with the effective coupling $\lambda_Q \propto g_{\mathrm{YM}}^2\Delta$ held fixed. The relevant heavy-heavy-light-light four-point correlators are known to all orders in $\lambda_Q$ in this limit. For certain classes of operators, the perturbative series can be resummed to obtain exact finite-coupling expressions. We use these results to determine the corresponding EEC, whose dynamical part can be extracted from a double discontinuity of the four-point correlator. We determine the EEC at finite effective coupling and study its physical
behavior as a function of $\lambda_Q$ and the angular separation between
the detectors, both at generic angles and in the collinear and back-to-back
limits.
}

\begin{document}
\maketitle

\section{Introduction}
\label{sec:introduction}

Energy correlators characterize the angular distribution of energy produced in a quantum field theory. One of the simplest correlations between distinct detectors is the energy–energy correlator (EEC), obtained by inserting two calorimetric detectors at future null infinity. For a state $|\Psi\rangle$, it is defined by \begin{equation}
  \operatorname{EEC}(\mathbf n_1,\mathbf n_2)
  \equiv
  \frac{\langle \Psi|\mathcal E(\mathbf n_1)\mathcal E(\mathbf n_2)|\Psi\rangle}
       {\langle\Psi|\Psi\rangle}\, ,
  \label{eq:EEC-intro}
\end{equation}
where the energy-flow operator $\mathcal E(\mathbf n)$ measures the energy arriving at infinity in the direction $\mathbf n$. In particular, $\mathcal E(\mathbf n)$ is a light-ray operator and hence constructed by integrating the appropriate component of the stress tensor along a null direction. 
Energy correlators were introduced as infrared-safe event-shape observables in gauge theory
\cite{Sterman:1975xv,Basham:1978bw,Basham:1978zq}
and were later developed further in the context of conformal collider physics
\cite{Hofman:2008ar}. They sum inclusively over final states, while retaining angular information that probes the underlying dynamics.

The collinear and back-to-back regions probe distinct
universal limits. At small angular separation, products of energy-flow operators are organized by the light-ray OPE, whereas the back-to-back limit is governed by Sudakov dynamics
\cite{Kravchuk:2018htv,Kologlu:2019mfz,Chang:2020qpj,
Dixon:2019uzg,Korchemsky:2019nzm, Chen:2026lic}.
These features have made the EEC an important observable in both fundamental studies of QFT and collider phenomenology; see Ref.~\cite{Moult:2025nhu} for a recent comprehensive review.

The EEC is particularly tractable in CFTs because it can be extracted from an appropriate Lorentzian continuation and detector limit of a Euclidean correlation function. The relevant discontinuities select the physical Wightman ordering, relating event shapes to the extensive toolkit developed for studying four-point functions
\cite{Belitsky:2013ofa,Belitsky:2013xxa,Belitsky:2013bja}.
This relation is especially effective in four-dimensional $\mathcal N=4$ super Yang--Mills (SYM), where supersymmetry, integrability, supersymmetric localization, and holography provide access to correlation functions over a wide range of coupling. 

A broad range of complementary techniques have been used in the study of energy correlators in $\mathcal N=4$ SYM. Their perturbative expansion is known to
high loop order
\cite{Belitsky:2013ofa,Belitsky:2013xxa,Belitsky:2013bja, Henn:2019gkr,Dixon:2019uzg},
while the light-ray OPE provides a nonperturbative description of their angular limits
\cite{Kologlu:2019mfz,Korchemsky:2019nzm}.
At strong coupling, holography and the AdS curvature expansion \cite{Alday:2023jdk, Alday:2023mvu, Fardelli:2023fyq} give systematic
access to supergravity and stringy corrections
\cite{Hofman:2008ar,Goncalves:2014ffa,Dempsey:2025yiv,Ren:2026zxs,Jackson:2026eml},
complemented by numerical bootstrap bounds at finite coupling
\cite{Dempsey:2025yiv}.
Energy correlators in $\mathcal N=4$ SYM thus offer a controlled setting for exploring how energy distributions evolve across coupling regimes.

Most studies of conformal event shapes consider states created by operators of fixed dimension. In
the present work the source is instead a half-BPS operator with parametrically large-$R$ charge (or equivalently large conformal dimension).\footnote{Earlier work~\cite{Chicherin:2023gxt} studied energy correlators in heavy states in the conventional planar limit, rather than the large-charge 't Hooft limit considered here. Our study is partially motivated by the absence of an all-order finite-coupling result available in the planar limit.} This introduces a macroscopic scale into an otherwise scale-invariant problem and asks a qualitatively different question: how does energy propagate when the state itself produces a nontrivial semiclassical background? The study of heavy sources also has a broader motivation. Energy-energy correlators are increasingly used to study how a nontrivial medium modifies the angular distribution of energy within jets, particularly in the quark-gluon plasma
\cite{Andres:2023xwr,Barata:2023bhh}. The large-charge half-BPS states in $\mathcal{N}=4$ SYM considered here are nonthermal and are not intended as models of heavy-ion matter. Nevertheless, they provide a controlled analytic laboratory for studying how a nontrivial macroscopic background modifies energy flow. Moreover, the large-charge 't Hooft limit, which we will review shortly, makes it possible to study these properties at finite coupling within an analytically controlled setting.

QFTs often simplify in sectors carrying a parametrically large quantum number. The large-charge expansion, initiated in its modern form
in~\cite{Hellerman:2015nra} and subsequently developed
in~\cite{Monin:2016jmo,Alvarez-Gaume:2016vff,
Jafferis:2017zna}, exploits this observation for operators carrying a large global charge. Supersymmetric CFTs with moduli spaces provide an especially rich setting for this program.  In these theories a large $R$-charge insertion can select a point, or more generally a trajectory, on the moduli space so that the semiclassical large-charge description is directly related to physics on the Coulomb branch. This connection has been explored extensively in four-dimensional $\mathcal N=2$ superconformal theories, where supersymmetric localization \cite{Pestun:2007rz} gives independent and often exact information about extremal correlators~\cite{Hellerman:2017sur,Hellerman:2018xpi,Grassi:2019txd}.  

In this work, we consider $\mathcal N=4$ SYM in the large-charge 't~Hooft limit, which combines large charge with weak Yang--Mills coupling. For a half-BPS operator of dimension $\Delta$, this limit is defined schematically by~\cite{Bourget:2018obm,Caetano:2023zwe}
\begin{equation}
  \Delta\rightarrow\infty,
  \qquad  {\rm with} \quad \lambda_Q \propto 
\Delta \, g_{\rm YM}^{2}
  \quad\text{fixed}\, .
  \label{eq:large-charge-thooft-intro}
\end{equation}
This limit is neither the ordinary planar 't~Hooft limit nor the pure large-charge limit at fixed Yang--Mills coupling.  The suppression of $g^2_{\rm YM}$ organizes perturbation theory, whereas combinatorial enhancements from the heavy operator compensate powers of the coupling, leaving a nontrivial function of $\lambda_Q$. A relevant class of observables to this work are heavy-heavy-light-light (HHLL) correlators,
\begin{equation}
    \big\langle
    \mathcal H(x_1)\,
    \mathcal O_{\rm L}(x_2)\,
    \mathcal O_{\rm L}(x_3)\,
    \overline{\mathcal H}(x_4)
    \big\rangle ,
\end{equation}
where $\mathcal{H}$ denotes the heavy operators whose scaling dimension we take to be large, while the dimensions of the light operators and the rank of the gauge group remain fixed. At leading order in the inverse charge, these correlators have been determined to all orders in $\lambda_Q$
\cite{Caetano:2023zwe,Brown:2024yvt,Brown:2025cbz}. Their perturbative
expansion is organized in terms of conformal ladder integrals and, for suitable
heavy operators, can be resummed to obtain exact finite-$\lambda_Q$ results.

We exploit these finite-coupling HHLL correlators to study the EEC in the
large-charge 't Hooft limit. The heavy operator $\mathcal H$ creates the state
$|\Psi\rangle$ appearing in \eqref{eq:EEC-intro}, while the light operator is
chosen to be the superconformal primary in the stress-tensor multiplet of $\mathcal N=4$ SYM. Through the
supersymmetric Ward identity, the light insertions can be converted into energy-flow operators
\cite{Belitsky:2013ofa,Belitsky:2013xxa,Belitsky:2013bja}.
We show that, in the large-charge limit and after the appropriate Lorentzian continuation, the resulting EEC
can be extracted from a double discontinuity of the HHLL correlator, thereby retaining its full dependence on the finite effective coupling.

Applying this representation, we determine the EEC analytically through three loops, with the heavy-source dependence encoded in colour factors multiplying universal angular functions. For canonical operators, we numerically evaluate the resummed detector transform at finite coupling and study how the angular distribution depends on the coupling and the massive spectrum of the Coulomb-branch background sourced by the heavy insertions. The leading collinear singularity is one-loop exact at finite effective coupling, whereas the regular back-to-back value receives higher-loop corrections. We determine the latter through four loops and derive its strong-coupling expansion. We further investigate the interplay between detector angle $\xi$ and coupling $\lambda_Q$ through the combination $\lambda_Q\xi$, which probes different interaction regimes as the angular resolution varies.

The rest of the paper is organised as follows. In section~\ref{sec:large-charge-thooft}, we review the large-charge 't Hooft limit and introduce the HHLL correlators, focusing on two classes of heavy operators for which the all-loop perturbative series can be resummed. In section~\ref{sec:dDISC} we show
how the EEC can be expressed as an integral over two discontinuities of the four-point correlator. We illustrate this construction with an explicit calculation of the one-loop result. In section~\ref{sec:large-charge-eec-results} we use the previous two sections to obtain explicit perturbative results for the  EEC in large-charge 't Hooft limit through three loops,
and evaluate the resummed EEC numerically at finite coupling. We also examine the  strong coupling
behaviour and further consider their property in terms of effective coupling $\lambda_Q \xi$, followed by a discussion of the collinear and back-to-back limits. We conclude in section~\ref{sec:con and out} with a discussion of our results and future directions.

The paper includes three appendices containing technical details. Appendix~\ref{app:angular-iterated-integrals} introduces angular iterated integrals and presents comparisons with known planar results, appendix~\ref{app:three-loop-coefficients} gives the complete three-loop ladder-product integral, and appendix~\ref{app:strong-expansion} derives the strong-coupling expansion at the back-to-back endpoint.

\vspace{0.5cm}

{\bf Note added.} While this work was being finalized, we became aware of concurrent work by Alba Grassi, Cristoforo Iossa, Luigi Tizzano and Alexander Zhiboedov~\cite{GITZ-to-appear}, which has some overlap with our results. 

\section{Review of the large-charge 't Hooft limit and HHLL correlators in \texorpdfstring{$\mathcal N=4$}{N=4} SYM}
\label{sec:large-charge-thooft}

The large-charge expansion provides a semiclassical reorganization of correlation functions in which the quantum number carried by a local operator is taken to be large.  A particularly useful version for the study of correlation function in gauge theories is the large-charge 't Hooft limit~\cite{Bourget:2018obm}.  For gauge group $SU(N)$, with $N$ held fixed, one considers half-BPS operators $\mathcal{O}_H$ whose dimension $\Delta_H \to\infty$, while the large-charge 't Hooft coupling $\lambda_Q=\Delta_H g_{\rm YM}^{2}/(16\pi^2)$ is held fixed.  This limit should be distinguished from both the usual planar 't Hooft limit and the pure large-charge limit at fixed $g_{\rm YM}^{2}$.  We will  consider the large-charge 't Hooft limit for HHLL correlators of half-BPS operators in $\mathcal{N}=4$ SYM.

\subsection{Exact results for HHLL correlators}

The half-BPS superconformal primaries in $\mathcal{N}=4$ SYM are
\begin{align} \label{T-def}
  T_{p_1, \ldots, p_n}(x,Y) &=  \frac{p_1 \cdots p_n} {p} T_{p_1}(x,Y) \ldots T_{p_n}(x,Y),  \cr
  \qquad T_p(x,Y) & =\left(\frac{\tau_2}{4\pi}\right)^{p/2}
  \operatorname{Tr}\!\left(\Phi^{I_1}(x)\cdots\Phi^{I_p}(x)\right)   Y_{I_1}\cdots Y_{I_p} \, ,
\end{align}
where $p=\sum_i p_i$ and $SO(6)_R$ vector $Y_I$ is null $Y \cdot Y =0$.  We are interested in four-point functions of the form,
\begin{align} \label{eq:def-reduce}
\frac{\langle \mathcal{H}(x_1, Y_1)   \mathcal{H}(x_2, Y_2)   \mathcal{O}_2(x_3, Y_3)   \mathcal{O}_2(x_4, Y_4)  \rangle}{\langle \mathcal{H}(x_1, Y_1)    \mathcal{H}(x_2, Y_2)  \rangle} = {\rm free \,\, part} + R_4(x_i, Y_i) {\mathcal T}_{\mathcal{H}}(u,v) \, , 
\end{align}
where cross ratios $u, v$ are defined as
\begin{align} \label{eq:uv-def}
    u = \frac{x_{12}^2 x_{34}^2}{x_{13}^2 x_{24}^2} \, , \qquad  v = \frac{x_{23}^2 x_{14}^2}{x_{13}^2 x_{24}^2} \, . 
\end{align}
Here $ \mathcal{O}_2=T_2$ is a single-trace operator as defined in \eqref{T-def}, and the operator $\mathcal{H}$ is a linear combination of $T_{p_1, \ldots, p_n}$ with dimension $\Delta_{\mathcal{H}}$, which ultimately will be taken to large in the large-charge 't Hooft limit. In the above expression the first term is determined by free-theory Wick contractions, and the structure of the remaining dynamical part is determined by the superconformal Ward identity \cite{Eden:2000bk, Nirschl:2004pa}. In particular, the prefactor $R_4(x_i, Y_i)$ is fixed by supersymmetry to be
\begin{align}
    R_4(x_i, Y_i) = \left(\frac{Y_3 \cdot Y_4}{4\pi^2 x_{34}^2} \right)^2 \frac{(z-\alpha)(z-\bar \alpha)(\bz-\alpha)(\bz-\bar \alpha) z\bz}{ \alpha^2 \bar  \alpha^2 (1-z)(1-\bz)} \, , 
\end{align}
with 
\begin{align}
    \alpha \bar \alpha &= \frac{Y_1 \cdot Y_2 Y_3 \cdot Y_4}{Y_1 {\cdot} Y_3 Y_2 \cdot Y_4} \, ,\qquad     (1-\alpha)(1- \bar \alpha) = \frac{Y_2 \cdot Y_3 Y_1 \cdot Y_4}{Y_1 \cdot Y_3 Y_2 \cdot Y_4} \, ,
    \end{align} 
    and 
    \begin{align}
    z\bz &= u \qquad  (1-z)(1-\bz) =v \, . 
\end{align}
Importantly, all dynamical information is contained within the reduced correlator $\mathcal T_{\mathcal H}(u,v)$, which will be our main interest.   

We will consider the large-dimension (or large-charge) limit where $\Delta_{\mathcal H}$ is the largest parameter of the system. In this limit it was shown in~\cite{Brown:2024yvt}, using the method of chiral Lagrangian insertion \cite{Eden:2012tu}, that the four-point function admits a 't Hooft like topological expansion. At leading order the perturbative loop expansion of ${\mathcal T}_{\mathcal H}(u,v)$ is given by products of ladder integrals only 
\begin{align} \label{eq:TH}
{\mathcal T}_{\mathcal H}(u,v) =  \frac{1}{u} \sum_{L=1}^{\infty} \sum_{\ell=0}^{L} d_{{\mathcal H}, L} (-\lambda_Q)^L 
P^{(\ell)}(u,v) P^{(L-\ell)}(u,v) \, , 
\end{align}
where we have introduced large-charge 't Hooft coupling  $\lambda_Q = \Delta_{\mathcal H} \, g_{\rm YM}^2/(16\pi^2)$, and $P^{(\ell)}(u,v)$ is the four-dimensional conformal ladder integral at $\ell$ loops. 
The analytic expression for these ladder integrals is known at all loop orders and is given by \cite{Usyukina:1993ch} 
\begin{equation}
 P^{(\ell)}(u,v)
 =- \frac{u}{z-\bz}
 \sum_{r=0}^{\ell}
 \frac{(-1)^{r}(2\ell-r)!}{r!(\ell-r)!\ell!}
 \log^r v\,
 \left[
 \operatorname{Li}_{2\ell-r}(1-z)
 -\operatorname{Li}_{2\ell-r}(1-\bz)
 \right]\, , 
 \label{eq:ladder def}
\end{equation}
for $\ell\geq1$, with $P^{(0)}=1$. 

The all-loop expression \eqref{eq:TH} is valid in the limit $\Delta_{\mathcal{H}} \to \infty$ with $\lambda_Q$ held fixed. The colour factor $d_{{\mathcal H}, L}$ depends on the precise form of the heavy operator ${\mathcal H}$ and the rank of gauge group. For certain special classes of heavy operators the colour factor simplifies and one may resum \eqref{eq:TH}. In such cases, we obtain a non-perturbative result for the HHLL correlators with finite $\lambda_Q$. In this paper, we consider two special classes of operators. The first, which we refer to as ``canonical operators,'' were introduced in~\cite{Brown:2024yvt, Brown:2025cbz} and are reviewed in section~\ref{sec:canonical-operators}. The second is the coherent state of~\cite{Aprile:2024lwy, Aprile:2025hlt}, whose properties in the large-charge 't Hooft limit we investigate in section~\ref{sec:coherent-states}.

\subsection{Canonical operators}
\label{sec:canonical-operators}

A family of ``canonical operators", denoted $\mathcal O_K^m$, were identified in~\cite{Brown:2025cbz} for SU$(N)$ $\mathcal N=4$ SYM, 
\begin{align} \label{eq:cn-const}
\mathcal O_K^m = T_{\underbrace{{}_{K, \ldots, K}}_m} + \sum_{\bold n} c_{\bold n} T_{\bold n} \, ,  
\end{align} 
where $T_{\bold n}$ are the operators that have the same dimension as $T_{K, \ldots, K}$, namely $\Delta = mK$, and due to trace relations $K$ cannot be larger than $N$.  The coefficients $c_{\bold n}$ are determined by a Gram-Schmidt diagonalization procedure as detailed in \cite{Brown:2025cbz}. Also discussed in~\cite{Brown:2025cbz}, canonical operators exhibit particularly simple properties in the large-charge 't Hooft limit, which we define by the limit $m\to\infty$ with $K$, $N$, and the effective coupling $\lambda_Q$ held fixed. For these operators, we define the coupling to be
\begin{align}
    \lambda_Q= \frac{ m K g_{\rm YM}^2 } {16\pi^2} \, .
\end{align}
  In particular, the correlation functions with two insertions of $\mathcal O_K^m$ at $x_1$ and $x_2$ source a Coulomb-branch saddle, with the scalar profile given by
\begin{equation}
  \Phi_{\rm cl}^{I}(x)=
  \frac{2\sqrt{\lambda_Q}}{\sqrt{K d_{12}}}
  \left[
    e^{i\theta}\Omega_K^{(N)}\frac{Y_1^I}{(x-x_1)^2}
    +e^{-i\theta}\Omega_K^{(N)}\frac{Y_2^I}{(x-x_2)^2}
  \right],
  \label{eq:canonical-saddle}
\end{equation}
where $d_{ij}=Y_i \cdot Y_j /x_{ij}^2$ and $\theta$ is a compact saddle modulus and
\begin{equation}
  \Omega_K^{(N)}=\diag\!\left(e^{2\pi i/K},e^{4\pi i/K},\ldots,e^{2\pi i},0,\ldots,0\right).
\end{equation}
The scalar profile distribution breaks the gauge group according to
\begin{equation}
  SU(N)\longrightarrow U(1)^{K-1}\times U(N-K)\, .
  \label{eq:symmetry-breaking}
\end{equation}
Consequently $(K-1)+(N-K)^2$ multiplets remain massless, whereas the off-diagonal modes acquire position-dependent masses whose dimensionless coefficients are
\begin{align}
M_s&= \frac{2\sqrt{\lambda_Q}}{\sqrt{K}}\, \sin\frac{\pi s}{K},
  &&{\rm for} \quad s=1,\ldots,K-1 \, ,
 \qquad \text{with multiplicity }K,
  \label{eq:mass-spectrum-a}\\
  M_s&=\frac{\sqrt{\lambda_Q}}{\sqrt{K}} \, , && {\rm for} \quad s=K\, , 
 \qquad \qquad \text{with multiplicity }2K(N-K) \, .
  \label{eq:mass-spectrum-b}
\end{align}

Explicit examples of canonical operator have been constructed in \cite{Brown:2025huy} using the procedure briefly outlined in \eqref{eq:cn-const} and the discussion afterwards. For example in the SU$(2)$ case, the trace identities imply that the only possible operator is $\mathcal O_2^m = T_{2, \ldots, 2}:=T_2^m$, which is the case considered in \cite{Caetano:2023zwe}. For $N>2$, the construction of canonical operator becomes highly non-trivial, and requires $K \geq 3$ \cite{Brown:2025cbz}.\footnote{The $K=2$ case was argued to describe the large-$N$ limit of the so-called dual giant gravitons with dimension $\beta N$ in the limit $\beta \to \infty$ and $\lambda_Q = \beta N g^2_{\rm YM}/(16\pi^2)$ being fixed~\cite{Brown:2025huy}.}  The first non-trivial case is SU$(3)$, and it was found to be \cite{Brown:2024yvt, Grassi:2024bwl, Brown:2025cbz} 
\begin{align} 
\mathcal O_3^m = T_{3}^m + \sum_{p, q} c_{p, q} \, T_{3}^p \, T_2^q \, , 
\end{align} 
with $3p+2q = 3m$ and 
\begin{equation}
c_{p, q} = 
\frac{(-1)^{q}\, (p+1)_{q/3}}
     {24^{\,q/3}\bigl(q/3\bigr)!}\, .
\end{equation}
We refer further examples for more general $K$ and $N$ to the reference \cite{Brown:2025cbz}. Below we will consider the correlation functions involving these operators.  

 In this limit, the $(n{+}2)$-point correlation functions with two  canonical operators $\mathcal O_K^m$ and $n$ other light operators are effectively described by $n$-point function of  light operators only in the non-trivial background \eqref{eq:canonical-saddle}. We will be interested in the HHLL correlators with two light operators being $\mathcal O_2 \equiv T_2$:  
 \begin{align}
 \langle \mathcal O_K^m (x_1, Y_1) \, \mathcal O_K^m (x_2, Y_2) \, \mathcal O_2 (x_3, Y_3) \, \mathcal O_2 (x_4, Y_4) \rangle \, ,
 \end{align} 
 normalised by the two-point function $ \langle \mathcal O_K^m (x_1, Y_1) \, \mathcal O_K^m (x_2, Y_2) \rangle$. 
In the large-charge 't Hooft limit, it  was argued in~\cite{Brown:2025cbz} that the reduced correlator is given by
\begin{equation}
  \mathcal T_{m,K}(u,v;\lambda_Q)
  \simeq \frac{1}{2u}\sum_s\left[t_s(u,v;\lambda_Q)^2-1\right],
  \qquad
  t_s(u,v;\lambda_Q)=\sum_{\ell=0}^{\infty}(-M_s^2)^\ell
  P^{(\ell)}(u,v)\, ,
  \label{eq:hhll-master}
\end{equation}
where $M_s$ are given in \eqref{eq:mass-spectrum-a} and \eqref{eq:mass-spectrum-b}. The sum includes their multiplicities, so that
\begin{equation}
 d_{\mathcal H,L}=\frac{1}{K^L}\left[
 \frac K2\sum_{s=1}^{K-1}\left(4\sin^2\frac{\pi s}{K}\right)^L
 +K(N-K)\right] \, . 
 \label{eq:canonical-colour-factor}
\end{equation}
 Physically, the ladder sum is the perturbative expansion of two massive scalar propagators between the light insertions in the inhomogeneous Coulomb-branch background \cite{Giombi:2020enj}.   $ \mathcal T_{m,K}(u,v;\lambda_Q)$ has the generic form of the HHLL correlator in the large-charge 't Hooft limit as in \eqref{eq:TH}, with simplified expression of the colour factor. 

Remarkably, the perturbative series in \eqref{eq:hhll-master} admits a closed resummation in this case~\cite{Broadhurst:2010ds}, and one obtains the expression for HHLL correlator that is valid at finite coupling. Indeed, the convergent perturbative series admits the resummed representation
\begin{equation}
 t_s(u,v;\lambda_Q)
 =\frac{u}{\sqrt v}
 \sum_{r=1}^{\infty}
 \frac{
 r\,e^{-\sigma\sqrt{r^2+4M_s^2}}\sin(r\phi)}
 {\sqrt{r^2+4M_s^2}\sin\phi}\, ,
 \label{eq:ts-spectral}
\end{equation}
where for $u=z\bz$ and $v=(1-z)(1-\bz)$ we have defined
\begin{equation}
 e^{-\sigma}=\sqrt{(1-z)(1-\bz)},
 \qquad
 e^{i\phi}=\sqrt{\frac{1-z}{1-\bz}}.
\end{equation}
This representation defines the HHLL correlator at finite $\lambda_Q$. At large $\lambda_Q$, it may instead be reorganized as a sum of worldline-instanton contributions in terms of modified Bessel functions \cite{Caetano:2023zwe},  
\begin{align}
t_s(u, v; \lambda_Q) = \frac{u}{\sqrt{v}} \sum_{n \geq 0} \left[ W_s(\phi + 2\pi n,\sigma)+W_s(-\phi+2\pi(n+1),\sigma) \right] 
\label{eq:worldline-instanton-rep}
\end{align}
with 
\begin{align}
W_s(x, \sigma) = \frac{2 M_s x K_1 (2 M_s \sqrt{x^2 + \sigma^2})}{\sin(x) \sqrt{x^2 + \sigma^2}} \, ,
\end{align}
and $K_1 (x)$ is the modified Bessel function. Terms larger in $n$ are increasingly exponentially suppressed at strong coupling.

\subsection{Coherent states in the large-charge 't Hooft limit }
\label{sec:coherent-states}

Another type of heavy operator we consider is the coherent state
\cite{Aprile:2024lwy, Aprile:2025hlt}
\begin{align}
\mathcal O_{{\rm ch},\, \alpha}
=
\exp\!\left(\frac{\alpha}{\sqrt{2}}\mathcal O_2\right) \, ,
\end{align}
where we have used the normalisation
$\langle\mathcal O_2\, \mathcal O_2\rangle=2(N^2-1)$, and we will consider the HHLL correlator of the following form, 
\begin{align} \label{eq:OcO2}
\langle \mathcal O_{{\rm ch},\, \alpha}(x_1, Y_1) \mathcal O_{{\rm ch},\, \alpha}(x_2, Y_2)  \mathcal O_2(x_3 ,Y_3) \mathcal O_2(x_4 ,Y_4)  \rangle \, . 
\end{align}
Although this state has no definite scaling dimension, its mean is $\langle\Delta_{\mathcal O_{\rm ch}}\rangle=(N^2-1)\rho$, where
$\rho=2|\alpha|^2/(1-2|\alpha|^2)$.
Thus $\rho\gg1$ gives $\langle \Delta_{\mathcal{O}_{\rm ch}} \rangle\gg N^2$, while relative charge fluctuations are suppressed at large $N$. These properties once again allow us to define the large-charge 't Hooft expansion, in which we take the large-$N$ limit and then setting $\rho\to\infty$ while keeping large-charge 't Hooft coupling being fixed; in this case, it is given as
\begin{align} \label{eq:coherent-coupling}
    \lambda_{\rm ch}=\frac{N\rho g_{\rm YM}^2}{16\pi^2}
=\frac{N}{N^2{-}1}
\frac{\langle\Delta_{\mathcal O_{\rm ch}}\rangle g_{\rm YM}^2}{16\pi^2}\, . 
\end{align} 
This additional factor is due to the large-$N$ colour factors compensate part of the enhancement by the total charge of the operator. This definition is also consistent with the explicit results of integrated correlator \cite{Aprile:2026urq} by applying results of \cite{Paul:2022piq, Paul:2023rka, Brown:2023why} for $\langle (\mathcal{O}_2)^p (\mathcal{O}_2)^p \mathcal{O}_2\mathcal{O}_2 \rangle$ computed by supersymmetric localisation \cite{Binder:2019jwn}. Expressed the results in the large-charge ’t Hooft expansion discussed above,  the result of \cite{Aprile:2026urq} of integrated correlator for the coherent state HHLL correlator has a well-defined all order expansion
\begin{align} \label{eq:integratedcor}
\mathcal C_{{\rm ch}} (\lambda_{\rm ch}) 
=\sum_{L=1}^{\infty}
(-1)^{L+1}
\frac{(2L+1)^2}{(L+1)^2(L+2)}
\binom{2L}{L}^{3}
\zeta(2L+1) \, 
\lambda_{\rm ch}^L  \, ,
\end{align}
where we have omitted terms that are higher order in $1/\rho$ and $1/N$. 

In this limit, the reduced HHLL correlator admits a perturbative expansion in products of ladder integrals as in \eqref{eq:TH}, with $\lambda_Q$ replaced by $\lambda_{\rm ch}$, following the same argument \cite{Brown:2024yvt}. With this structure, the colour factors $d_{\mathcal O_{\rm ch},L}$ are determined by the known integrated correlator \cite{Binder:2019jwn} as done in \cite{Brown:2024yvt} for generic heavy operators.  The integrated correlator is defined as \cite{Binder:2019jwn}
\begin{align}
\mathcal C_{{\rm ch};\, N} (g_{\rm YM}^2) = -\frac{2}{\pi} \int_0^{\infty} dr \int^{\pi}_0 d \theta \frac{r^3 \sin^2(\theta)}{u\, v} \mathcal{T}_{\rm ch}(u, v) \, , 
\end{align}
where $\mathcal{T}_{\rm ch}(u, v)$ is the reduced correlator of the four-point function \eqref{eq:OcO2}. Using the integrated correlator result \eqref{eq:integratedcor} and the integrated result for the product of ladder integrals \cite{Dorigoni:2021guq},
\begin{align}
     -\frac{2}{\pi} \int_0^{\infty} dr \int^{\pi}_0 d \theta \frac{r^3 \sin^2(\theta)}{u^2\, v} P^{(\ell)}(u,v) P^{(L-\ell)}(u,v) = -\frac{1}{2} \binom{2L{+}2}{L+1} \zeta(2L+1) \, ,  
\end{align}
we find the coherent-state colour factor in the large-charge 't Hooft limit is given by 
\begin{align} \label{eq:coherent-color}
d_{\mathcal O_{\rm ch},L} = \frac{2L+1}{(L+1)^2(L+2)}
\binom{2L}{L}^{2}  \, . 
\end{align}
These colour factors can be generated by an integral transform of the resummed ladder contribution appearing in the canonical-state correlator. We then find following representation for the coherent-state HHLL correlator, 
\begin{align} \label{eq:Tch}
\mathcal T_{\rm ch}(u,v;\lambda_{\rm ch})
&=\frac1u\int_0^1dr \, w(r)
\left[
\left(\mathcal B\!\left(u,v;16\lambda_{\rm ch}r^2\right) \right)^2-1
\right],
\end{align}
where 
\begin{equation} \label{eq:weight}
w(r)=
\frac{2}{\pi}(1-r^2)^2\,
{}_2F_1\!\left(
\frac12,\frac32;3;1-r^2
\right)
\end{equation}
and
\begin{align} \label{eq:SB}
  \mathcal B(u,v;x)
=
\sum_{\ell=0}^{\infty}(-x)^\ell P^{(\ell)}(u,v) \, . 
\end{align}
The square in \eqref{eq:SB} generates the ladder convolution and the subtraction removes the free contribution. Therefore, the result \eqref{eq:Tch} provides an exact result for HHLL of coherent state operator. 

\section{Energy-energy correlators as double discontinuities} \label{sec:dDISC}

In this section we will derive the large-charge EEC directly in position space, inspired by the discontinuity construction of~\cite{Belitsky:2013xxa}. In the large-charge 't Hooft limit we will show that the EEC can be expressed as an integral over two discontinuities, in contrast to the triple discontinuity expression found in~\cite{Henn:2019gkr} for the EEC of operators with finite dimensions.

\subsection{The position-space detector transform}
For a null vector $n^\mu = (1, \bm n)$, $|\bm n|=1$, the energy flow detector operator is given by 
\begin{equation}
    \mathcal E(n) = \int_{-\infty}^{\infty}\mathrm du \, \lim_{r\to\infty} r^2 \, n_i \, T^{0i}(r+u,r\mathbf n)\, ,
\end{equation}
where $T^{\mu \nu}$ is the stress-energy tensor.
In the source rest frame $q^\mu=(Q,\mathbf 0)$ we define the EEC associated with half-BPS sources $\mathcal O$ to be
\begin{equation}
\operatorname{EEC}_{\mathcal O}(\xi)
=
\frac{16\pi^2}{Q^2\,\sigma_{\mathcal O}(q)}
\int \mathrm d^4x\,e^{iq\cdot x}
\left\langle
\mathcal O^\dagger(x)\,\mathcal E(n_1)\,\mathcal E(n_2)\,\mathcal O(0)
\right\rangle \, , \label{eq:def-EEC}
\end{equation}
where $n_1,n_2$ are future-directed null vectors, $\mathcal O$ is a local operator of scaling dimension $\Delta_{\mathcal O}$, which sources the energy flow, and the angular variable is
\begin{align} \label{eq:angle-def}
    \xi = \frac12 (1- \bm n_1 {\cdot} \bm n_2) = \frac12  (1 -\cos(\theta)) \, . 
\end{align}
 Note that the correlators are Wightman correlators, and there is an overall normalisation by the total cross-section
\begin{equation}
\sigma_{\mathcal O}(q)=
\int \mathrm d^4x\,e^{iq\cdot x}
\left\langle \mathcal O^\dagger(x) \, \mathcal O(0)\right\rangle\, .
\end{equation}
In $\mathcal N=4$ SYM, the half-BPS operator $\mathcal O_2$ appearing in the
HHLL correlator belongs to the stress-tensor supermultiplet. Consequently, the
EEC defined in \eqref{eq:def-EEC} is determined by the four-point function
introduced in \eqref{eq:def-reduce}
\cite{Belitsky:2013xxa,Belitsky:2013bja}. We focus on the dynamical part of the
energy-energy correlator,  $\EEC_{\mathcal O}^{\rm dyn} \equiv \EEC_{\mathcal O}-\EEC_{\mathcal O}^{\rm free}$ which is encoded in the reduced correlator\footnote{In the following, we suppress the superscript ``$\mathrm{dyn}$'',  with the
understanding that $\EEC_{\mathcal O}$ denotes only the dynamical part of the EEC unless
stated otherwise.}.

To relate this four-point function to the EEC, we require the ordering HLLH, with the heavy operators inserted at positions $1$ and $4$. Starting
from the HHLL ordering used in \eqref{eq:def-reduce}, we exchange \(x_2\leftrightarrow x_4\) which acts as \((u,v)\mapsto(v,u)\). For the EEC we define new cross-ratios adapted to the HLLH ordering, namely
\begin{equation}
U=\frac{x_{12}^{2}x_{34}^{2}}{x_{14}^{2}x_{23}^{2}} = w \bw,
\qquad
V=\frac{x_{13}^{2}x_{24}^{2}}{x_{14}^{2}x_{23}^{2}} = (1-w)(1-\bw),
\label{eq:detector-uv}
\end{equation}
where the positions refer to the relabelled configuration. The original HHLL cross-ratios are therefore expressed in terms of the detector variables as $u=1/V$, $v=U/V$, and the reduced correlator is
\begin{equation}
    \Phi_{\mathcal H}(U,V) = \frac{1}{U}\mathcal T_{\mathcal H}\!\left(\frac{1}{U},\frac{V}{U}\right)
    = \frac{1}{V}\mathcal T_{\mathcal H}\!\left(\frac{1}{V},\frac{U}{V}\right) \, , 
    \label{eq:ps-crossing}
\end{equation}
where the second equality follows by exchanging the two identical light operators. Given this crossed correlator, we can find an expression for the EEC by combining arguments laid out in~\cite[section 2.4]{Belitsky:2013xxa} and~\cite[appendix C]{Chicherin:2023gxt}. 

We begin by analytically continuing the Euclidean four-point function to obtain the corresponding Wightman function. The dynamical dependence of the EEC can be encoded in a function $\mathcal G_{\mathcal O}(\gamma)$ of the single dimensionless variable
\begin{equation}
 \gamma(x)=
 \frac{2(x\cdot n_1)(x\cdot n_2)}
 {x^2(n_1\cdot n_2)}.
 \label{eq:ps-ward-op}
\end{equation}
The function $\mathcal G_{\mathcal O}(\gamma)$ is completely determined by the scalar reduced correlator, where we can use the Ward identity to convert the scalar correlator into the energy-flow correlator by acting with the fourth-order differential operator~\cite{Chicherin:2023gxt}
\begin{equation}
 \mathcal D_\gamma
 =\partial_\gamma^2
 \bigl[\gamma^2(1-\gamma)^2\partial_\gamma^2\bigr].
\end{equation}
Fourier transforming the source separation $x$ projects onto a state of momentum $q$. Dividing by the source two-point function then gives, in our normalization, the dynamical EEC
\begin{equation}
 \EEC_{\mathcal O}(\xi)
 =-\frac{8\pi^3}
 {({\Delta_{\mathcal O}}-1)\xi^3 S_{\Delta_{\mathcal O}+1}(q)}
 \int\d^4x\,
 \frac{e^{iq\cdot x}}
 {(-x^2+i0x^0)^{\Delta_{\mathcal O}+1}}\,
 \mathcal D_\gamma\mathcal G_{\mathcal O}(\gamma(x)) \, .
 \label{eq:ps-exact}
\end{equation}
For future-directed timelike $q$, the scalar Fourier integral is
\begin{equation}
 S_p(q):=\int\d^4x\,
 \frac{e^{iq\cdot x}}{(-x^2+i0x^0)^p}
 =\frac{2\pi^3(q^2/4)^{p-2}}
 {\Gamma(p)\Gamma(p-1)},
 \qquad
 \frac{S_{p+1}(q)}{S_p(q)}
 =\frac{q^2}{4p(p-1)}.
 \label{eq:ps-S}
\end{equation}
The second identity allows the source normalization to be written in the form used in~\eqref{eq:ps-exact}.

Finally, $\mathcal G_{\mathcal O}$ can be obtained directly from reduced correlator $\Phi_{\mathcal O}(U,V)$. The position space identity of~\cite[eq.~(2.39)]{Belitsky:2013xxa} allows us expresses this function as an integral over two discontinuities of the correlator. Using the shorthand $\Phi_{\mathcal O}(w,\bw)$ to mean $\Phi_{\mathcal O}(U,V)$ evaluated at $U=w\bw$, $V=(1-w)(1-\bw)$, we have
\begin{equation}
\mathcal G_{\mathcal O}(\gamma)
=-\frac{1}{16\pi^5}
\int_{-\infty}^0\d w\int_1^\infty\d\bw\,
\Disc_w\Disc_{\bw}\Phi_{\mathcal O}(w,\bw)\,
\mathcal K(w,\bw;\gamma)\, ,
\label{eq:ps-G}
\end{equation}
where
\begin{align}
\mathcal K(w,\bw;\gamma)
={}&
\frac{w-\bw}{w\bw(1-w)(1-\bw)}
\Bigg[
\frac{1}{\sqrt{(-1+w+\bw+\gamma)^2-4\gamma w\bw}}
\nonumber\\
&\hspace{2.2cm}
+\frac{1}{w\bw-\gamma}
+\frac{1}{(1-w)(1-\bw)-\gamma}
+\frac1\gamma
\Bigg] \, , 
\label{eq:ps-kernel}
\end{align}
and we are using the convention $\Disc_w f(w)=\frac{1}{2i}(f(w+i0)-f(w-i0))$. The two discontinuities are taken over $w<0$ and $\bw>1$, as implied by the integration domain.

\subsection{The large-charge EEC as a double discontinuity}
We can now take the large-charge limit of \eqref{eq:ps-exact}, which localises the integral at a saddle for which $\gamma=1/\xi$.

For simplicity we will again work in the rest frame $q=(Q,\bm0)$ and set $b=\Delta_{\mathcal H}+1$. We then deform the integration contour into the lower half-plane as
\begin{equation}
 x^0 = \frac{2b}{Q}(y^0-i), \quad \bm x = \frac{2b}{Q} \bm y. 
 \label{eq:ps-contour}
\end{equation}
Apart from constants independent of $y$, the Fourier weight then becomes
\begin{equation}
 \frac{e^{iq\cdot x}}{(-x^2+i0x^0)^b} \sim \exp \left({b[2iy_0-\log H(y)]}\right),
 \label{eq:ps-weight}
\end{equation}
with $H(y)=1+|\bm y|^2-y_0^2+2iy_0$.
It follows that the Fourier transform localises at
\begin{equation}
 x_*^\mu=-\frac{2ib}{q^2}q^\mu
 \label{eq:ps-saddle}
\end{equation}
with width $b^{-1/2}$ in the rescaled coordinates. Note that $\gamma(x_*)=1/\xi$. Consequently, for a well-behaved\footnote{By well-behaved we precisely mean holomorphic near the saddle and polynomially bounded along the deformed contour, with no singularities obstructing the contour deformation.} function $f$ independent of $\Delta_{\mathcal H}$,
\begin{equation}
 \int\d^4x\, \frac{e^{iqx}}{(-x^2+i0x^0)^b}\, f(\gamma(x)) = S_b\bigl[f(1/\xi)+O(b^{-1})\bigr] \, , 
 \label{eq:ps-localization}
\end{equation}
where the function $S_b$ is defined in \eqref{eq:ps-S}. 

Now suppose that the correlator $\Phi_{\mathcal O}(w,\bw)$ becomes $\Phi_{\mathcal H}(w, \bw)$ in the large-charge 't Hooft limit, and denote its transform by $\mathcal G_{\mathcal H}$. Inserting~\eqref{eq:ps-localization} into~\eqref{eq:ps-exact} in the large charge 't Hooft limit gives
\begin{equation}
 \EEC_{\mathcal H}(\xi) := \lim_{\Delta_{\mathcal O}\to\infty} \Delta_{\mathcal O} \EEC_{\mathcal O}(\xi)
 =-\frac{8\pi^3}{\xi^{3}}
       (\mathcal D_\gamma \,\mathcal G_{\mathcal H})(1/\xi).
 \label{eq:ps-general-limit}
\end{equation}
The large-charge energy correlator $\EEC_\mathcal{H}$ is then neatly expressed by defining the functional
\begin{equation}
 I[F](\xi)=2\int_{-\infty}^0\d w\int_1^\infty\d\bw\,
 \Disc_w\Disc_{\bw}F\bigl(w\bw,(1-w)(1-\bw)\bigr)
 \mathcal K(w,\bw;\xi),
 \label{eq:ps-I}
\end{equation}
where the kernel $\mathcal K$ is the same as in \eqref{eq:ps-kernel} with $\gamma = 1/\xi$. Since $I[\Phi_{\mathcal H}](\xi)=-32\pi^5\mathcal G_{\mathcal H}(1/\xi)$, we obtain
\begin{equation}
    \EEC_{\mathcal H}(\xi) = \frac{\mathbb D_\xi I[\Phi_{\mathcal H}](\xi)}{4\pi^2},
 \label{eq:ps-universal}
\end{equation}
where we have used chain rule to simplify $\xi^{-3}(\mathcal D_\gamma f)(1/\xi)=\mathbb D_\xi f(1/\xi)$, introducing a new differential operator
\begin{equation}
 \mathbb D_\xi=\frac1\xi\frac{\d}{\d\xi}\,
 \xi^2\frac{\d}{\d\xi}\,
 \frac{(1-\xi)^2}{\xi^2}\frac{\d}{\d\xi}\,
 \xi^2\frac{\d}{\d\xi}.
 \label{eq:ps-D}
\end{equation}
Equation \eqref{eq:ps-universal} expresses the EEC in the large-charge 't Hooft limit as an integral over two discontinuities of the correlator. Unlike the finite charge formula~\cite{Belitsky:2013xxa, Henn:2019gkr}, no additional discontinuity in $\gamma$ is required.

For the purpose of studying the EEC for the canonical operators, which contains different massive modes, it will also be convenient to define the single-mode function
\begin{equation}
\eec_{K,N;s}(\xi;\lambda_Q) := \frac{\mathbb D_\xi I[\left(t_s(1/V,U/V)\right)^2-1](\xi;\lambda_Q)}{8\pi^2}\, ,
\label{eq:ps-mode-decomposition}
\end{equation}
where the apparent extra factor of 2 in the denominator relative to \eqref{eq:ps-universal} comes from the definition of $\mathcal T_{\mathcal H}$ in \eqref{eq:hhll-master}. The EEC  for the canonical operator $\mathcal{O}_{K}^m$ with SU$(N)$ gauge group in the large-charge 't Hooft limit is then given by a sum over all the mass modes with their corresponding multiplicities
\begin{equation}
\EEC_{K, N}(\xi;\lambda_Q) = K\sum_{s=1}^{K-1}\eec_{K,N;s}(\xi;\lambda_Q) + 2K(N-K)\eec_{K,N;K}(\xi;\lambda_Q)\, ,
 \label{eq:finite-coupling-mode-sum}
\end{equation}
where the summation is over all the massive modes as in \eqref{eq:mass-spectrum-a}  and \eqref{eq:mass-spectrum-b}.

\subsection{Example: One-loop EEC}
\label{sec:one-loop-check}

We illustrate the detector transform at one loop. Using \eqref{eq:TH}, at one loop, we have
\begin{align} 
{\mathcal T}_{\mathcal H}(u,v) =  - \frac{2}{u}\,  d_{{\mathcal H}, 1}\, \lambda_Q\,
P^{(1)}(u,v)  + O(\lambda_Q^2) \, ,
\label{eq:one-loop-correlator}
\end{align}
where $d_{\mathcal H,1}$ depends on the particular choice of heavy operator. Using the crossing relation \eqref{eq:ps-crossing} and the scalar-box identity
$P^{(1)}(1/U,V/U)=P^{(1)}(U,V)/U$, we obtain
\begin{equation}
\Phi_{\mathcal H}^{(1)}(w,\bw)
=-\frac{2d_{\mathcal H,1}}{w\bw}
P^{(1)}\bigl(w\bw,(1-w)(1-\bw)\bigr).
\label{eq:ps-box}
\end{equation}
Taking two discontinuities across $w<0$ and $\bw>1$ gives
\begin{equation}
\Disc_w\Disc_{\bw}\Phi_{\mathcal H}^{(1)}(w,\bw)
=\frac{2\pi^2d_{\mathcal H,1}}{\bw-w},
\label{eq:ps-box-disc}
\end{equation}
and substituting into \eqref{eq:ps-I} we find
\begin{equation}
I[\Phi_{\mathcal H}^{(1)}](\xi)
=8\pi^2d_{\mathcal H,1}\,
\xi\bigl[\Li_2(1-\xi)-\zeta_2\bigr].
\label{eq:ps-box-I}
\end{equation}
Therefore, after the detector transform, we arrive at
\begin{equation}
\EEC_{\mathcal H}^{(1)}(\xi)
=\frac{\mathbb D_\xi
I[\Phi_{\mathcal H}^{(1)}](\xi)}{4\pi^2} = d_{\mathcal H,1} \left[\frac{12}{\xi} + 8\Li_2(1-\xi) - 24\log\xi - 8\zeta_2 - 26\right],
\label{eq:ps-one-loop}
\end{equation}
which is valid in the range $0<\xi\leq1$. This is in agreement with~\cite[eq.~(4.27)]{Chicherin:2023gxt}.  Thus the one-loop angular dependence is universal for any operators, their dependence enters only through $d_{\mathcal H,1}$, which is simply $N$ at one loop. 

To include the collinear endpoint, we interpret the singular term as a plus distribution and include a contact term:
\begin{equation}
\EEC_{\mathcal H}^{(1)}(\xi)
=4d_{\mathcal H,1}
\left[
3\left[\frac1\xi\right]_+
+2\Li_2(1-\xi)
-6\log\xi
-2\zeta_2
-\frac{13}{2}
+\frac52\delta(\xi)
\right]\, .
\label{eq:ps-one-loop-distribution}
\end{equation}
Here
\begin{align}
\int_0^1\d\xi\,\left[\frac1\xi\right]_+f(\xi)
=\int_0^1\d\xi\,\frac{f(\xi)-f(0)}{\xi}\, .
\end{align}
This endpoint completion agrees with
\cite[eq.~(6.10)]{Chicherin:2023gxt} and satisfies the
energy--momentum sum rules \cite{Korchemsky:2019nzm}
\begin{equation}
\int_0^1\d\xi\,\EEC_{\mathcal H}^{(1)}(\xi)=0\,,
\qquad
\int_0^1\d\xi\,\xi\,\EEC_{\mathcal H}^{(1)}(\xi)=0\, .
\label{eq:ps-one-loop-sum-rules}
\end{equation}

\section{Energy-energy correlators in the large-charge 't Hooft limit}
\label{sec:large-charge-eec-results}

We now want to apply the detector transform of section~\ref{sec:dDISC} to the large-charge
correlators described in sections~\ref{sec:canonical-operators} and \ref{sec:coherent-states}. We first describe our results in perturbation theory and then using its resummed
form we study them at finite and strong coupling. The perturbative result is universal for any heavy operators;
the dependence on the precise form of the operators enters through the colour factors.

\subsection{Perturbation theory}
\label{sec:eec-perturbation-theory}

For studying the perturbative expansion of the EEC in the large-charge 't Hooft limit, it is convenient to introduce the following quantities:
\begin{equation}
 \mathcal F^{(L)}(\xi)=\frac{(-1)^L}{4\pi^2}
 I\!\left[\sum_{\ell=0}^L
 P^{(\ell)}(1/V,U/V)P^{(L-\ell)}(1/V,U/V)\right](\xi)\, ,
 \label{eq:results-perturbative-expansion}
\end{equation}
and 
\begin{align}  \label{eq:results-perturbative-expansion2}
    \eec^{(L)}(\xi)=\mathbb D_\xi\mathcal F^{(L)}(\xi) \, . 
\end{align}
Here $I$ is the double-discontinuity integral defined in \eqref{eq:ps-I} and $\mathbb D_\xi$ is given in \eqref{eq:ps-D}. We note $\mathcal F^{(L)}(\xi)$ (and $\eec^{(L)}(\xi)$) are universal that is independent of the precise form of the heavy operator. From the all-loop expression \eqref{eq:TH}, it is easy to see that EEC of a given heavy operator $\mathcal H$ is then given by
\begin{align}
    \EEC_{\mathcal H}(\xi;\lambda_Q)
 = \sum_{L=1}^{\infty} d_{\mathcal H,L}\, \lambda_Q^L\, \eec^{(L)}(\xi) \, . \label{eq:EECH}
\end{align}
The perturbative computation can be performed using \texttt{HyperInt} \cite{Panzer:2014caa} with some minor adjustment.
We find that it is useful to express the results in the real angular basis defined in
Appendix~\ref{app:angular-iterated-integrals}. All $\mathcal J_w$ below
have argument $\xi$, and the detector angle $\theta$ is
defined in \eqref{eq:angle-def}. At one loop we find
\begin{align}
 \mathcal F^{(1)}(\xi)&=8\xi\mathcal J_{TC}
 =2\xi\bigl[\Li_2(1-\xi)-\zeta_2\bigr]\, ,\\
 \eec^{(1)}(\xi)&=\frac{12}{\xi}
 +8\Li_2(1-\xi)-24\log\xi-8\zeta_2-26 \, ,
 \label{eq:results-one-loop}
\end{align}
which agrees with \eqref{eq:ps-one-loop} after using \eqref{eq:EECH}. 
At two loops the HHLL correlator in  the large-charge 't Hooft limit contains
$(P^{(1)})^2+2P^{(2)}$.\footnote{A similar heavy correlator has been studied before \cite{Chicherin:2023gxt} albeit with a different order of limits. First taking the large $N$ and then taking the operator dimension to be large. That result contains a
different ladder integral combination, and its EEC therefore differs beyond one loop. In Appendix~\ref{app:large-charge-two-loop} we present the two loop result using this angular basis.}
Writing $r=\sqrt{\xi(1-\xi)}$, and introducing
\begin{equation}
 Q(\theta)=\mathcal J_{EECC}+\mathcal J_{EETC}
             -\frac{\zeta_2}{4}\mathcal J_{EE}\, ,
 \label{eq:results-angular-seed}
\end{equation}
the two-loop scalar function takes the following form
\begin{equation}
\begin{aligned}
 \mathcal F^{(2)}(\xi)=\frac{2\xi}{(1-\xi)^2}\Bigl[&
 \frac{\pi^4\xi(7-25\xi)}{360}-8\xi\zeta_3
 +\frac{4\xi}{3}(6\xi\zeta_3+\pi^2)\mathcal J_C\\
 &+4(1-\xi)(6-5\xi)\zeta_3\mathcal J_T
 -\frac{4\pi^2\xi^2}{3}\mathcal J_{CC}
 +\frac{2\pi^2(1-\xi)(3\xi-4)}{3}\mathcal J_{TC}\\
 &-32\xi\mathcal J_{CTC}-32(1-\xi)\mathcal J_{TCC}
 +32\xi\mathcal J_{CCTC}-64\xi^2\mathcal J_{CTCC}\\
 &+16(1-\xi)(6-5\xi)\mathcal J_{TCTC}
 +32\bigl((2\xi-1)Q(\theta)
                   +2r\,\partial_\theta Q(\theta)\bigr)\Bigr]\, ,
\end{aligned}
 \label{eq:results-two-loop}
\end{equation}
where $\theta$ is defined as in \eqref{eq:angle-def}. Let us briefly discuss the result and show why we have decided to use the basis of \ref{app:angular-iterated-integrals}. In the variable $\xi$ the results containts both the ordinary harmonic polylogarithms and an extra complex square root letter $\tfrac{i \sqrt{\xi}}{\sqrt{1-\xi}}$. In this new angular integrals, the $C,T$ words form the part without square-root kernels and are
ordinary harmonic polylogarithms in $\xi$. The remaining dependence
is carried by the $E$ letter,
$d\xi/[2\sqrt{\xi(1-\xi)}]$. In the rationalizing variable
$t=\tan(\theta/4)$ of Appendix~\ref{app:angular-iterated-integrals},
this gives the additional kernel $dt/(1+t^2)$, whereas the $C,T$
sector contains $t\,dt/(1+t^2)$. It separates the conjugate
root-of-unity letters at $t=\pm i$ and extends the ordinary HPL sector. Its
square-root dependence is organized by one function $Q$ and its
derivative with respect to the detector angle. In particular,
$\partial_\theta Q=\tfrac12(\mathcal J_{ECC}+\mathcal J_{ETC}
-\zeta_2\mathcal J_E/4)$.

Similar calculation can be performed at higher loops. In particular, at three loops, the result takes 
the following general form
\begin{equation}
 \mathcal F^{(3)}(\xi)=\frac{4\xi}{(1-\xi)^2}
 \Bigl[P_3(\xi)-32\bigl(Q(\theta)-2r\,\partial_\theta Q(\theta)
              -2\xi A(\xi)-4(2\xi-1)B(\xi)\bigr)\Bigr].
 \label{eq:results-three-loop}
\end{equation}
The definitions of $A,B$ and the complete expression for $P_3$
are given in Appendix~\ref{app:three-loop-coefficients}. We see that $P_3$ contains only $C,T$ words, while $A,B$ are successive
angular primitives of $Q$. No new letter is required: the largest word
weight is six, and the apparent poles at $\xi=1$ cancel in the full
answer. 

Applying $\mathbb D_\xi$ to~\eqref{eq:results-two-loop} and
\eqref{eq:results-three-loop} we obtain $\eec^{(2)}$ and $\eec^{(3)}$, and \eqref{eq:EECH} gives the EEC at two and three loops after specifying the colour factors. For example, specifying to the canonical operators, the colour factor is then given by \eqref{eq:canonical-colour-factor}, whereas for the coherent state operator we have the colour factor as in \eqref{eq:coherent-color}. 
The colour factors for various other types of heavy operators may be found in \cite{Brown:2024yvt}. 

\subsection{Finite coupling}
\label{sec:eec-finite-coupling}

At finite coupling we evaluate the same transform with the resummed ladder integrals
\begin{equation}
 \mathcal F_{\mathcal H}(\xi;\lambda_Q)
 =\frac{I[\Phi_{\mathcal H}](\xi)}{4\pi^2}
 =\frac1{8\pi^2}\sum_s I[t_s(1/V,U/V;\lambda_Q)^2-1](\xi),
 \qquad \EEC_{\mathcal H}=\mathbb D_\xi\mathcal F_{\mathcal H}.
 \label{eq:finite-coupling-integrand}
\end{equation}
The summation over $s$ includes the mode multiplicities. We analytically continue the worldline-instanton
representation~\eqref{eq:worldline-instanton-rep}
to the detector integration region $w<0$, $\bar w>1$.
We then evaluate the double discontinuity
$\Disc_w\Disc_{\bar w}[t_s(1/V,U/V;\lambda_Q)^2-1]$
and perform the integrals in~\eqref{eq:ps-I} numerically after acting on the kernel with the differential operators. We check convergence by refining the integration grid and increasing the number of terms retained in the worldline-instanton sum. We begin with $N=K=2$ because there is a single nonzero mass,
$M_1^2=2\lambda_Q$, of multiplicity two, so that
$\EEC_{2,2}=2\eec_{2,2;1}$. Moreover we plot $\xi \EEC_{N,K}$ as it avoids the divergent collinear region and renders the numerics more efficient. 

\begin{figure}[htbp]
 \centering
 \includegraphics[width=.77\textwidth]{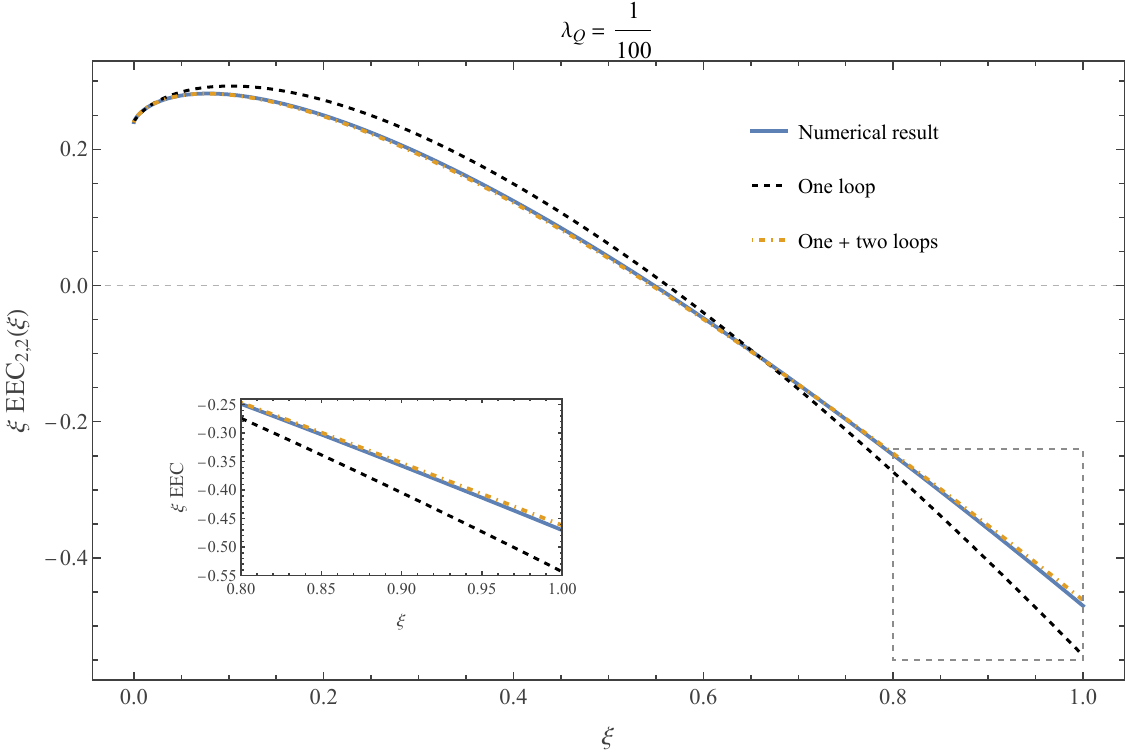}
 \caption{The $N=K=2$ result and its approximation through one and two loops at $\lambda_Q=1/100$. The curves share a collinear intercept at $\lim_{\xi\to0} \xi \EEC_{2,2} = 24\lambda_Q = 0.24$.}
 \label{fig:eec-small-coupling-check}
\end{figure}

\begin{figure}[htbp]
 \centering
 \includegraphics[width=.78\textwidth]{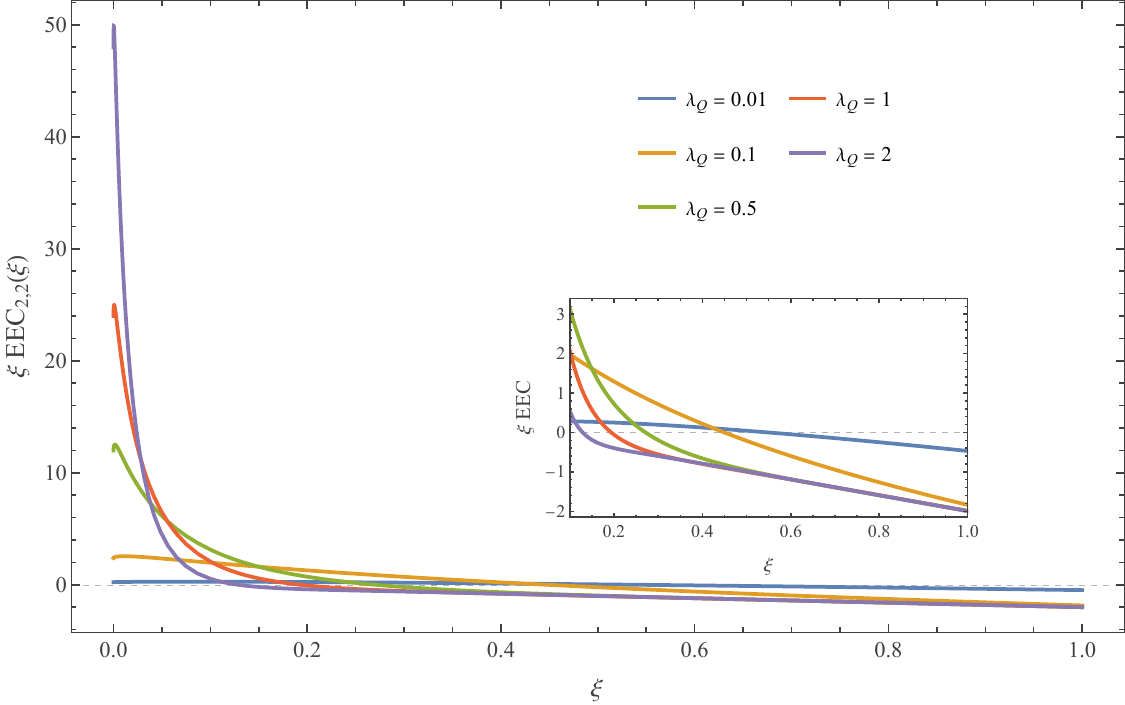}
 \caption{Angular dependence of the $N=K=2$ EEC at
 $\lambda_Q\in\{0.01,0.1,0.5,1,2\}$. The inset resolves
 the finite-angle behaviour for $\xi\geq0.1$.}
 \label{fig:eec-finite-coupling}
\end{figure}

\begin{figure}[htbp]
 \centering
 \includegraphics[width=.9\textwidth]{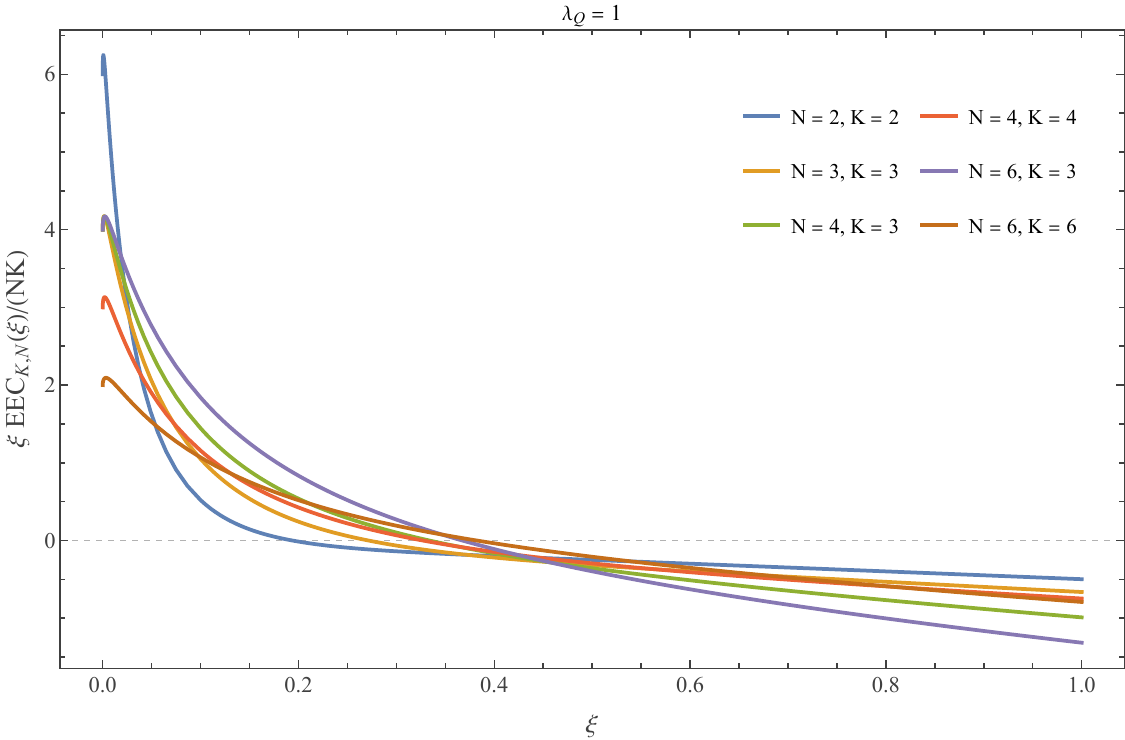}
 \caption{Canonical sources at $\lambda_Q=1$, showing
 $\xi\EEC_{K,N}/(NK)$. The collinear intercept is $12\lambda_Q/K=12/K$.}
 \label{fig:eec-finite-NK}
\end{figure}

We first test the calculation against perturbation theory. For
$N=K=2$ and $\lambda_Q=1/100$, the numerical curve for
$\xi\EEC_{2,2}(\xi)$ can be approximated already by considering just the first two loop contributions see figure \ref{fig:eec-small-coupling-check}.
Moving towards finite $\lambda_Q$, figure~\ref{fig:eec-finite-coupling} shows its
angular dependence at several values of the coupling. The endpoint
limits are discussed in section~\ref{sec:eec-endpoint-limits}.
For other values of $N,K$, the mass spectra in
\eqref{eq:mass-spectrum-a} and \eqref{eq:mass-spectrum-b} give
\begin{equation}
 \EEC_{K,N}(\xi;\lambda_Q)
 =K\sum_{s=1}^{K-1}\eec_{K,N;s}(\xi;\lambda_Q)
 +2K(N-K)\eec_{K,N;K}(\xi;\lambda_Q),
 \label{eq:results-finite-coupling-mode-sum}
\end{equation}
where $\eec_{K,N;K}$ denotes the mass branch associated with \eqref{eq:mass-spectrum-b}. For a range of values for $N$ and $K$ we notice the same angular pattern persists for a fixed value of coupling. That is, the positive collinear region is followed by a zero and a negative back-to-back region, with the spectrum only changing their relative size and the position of the zero.
Figure~\ref{fig:eec-finite-NK} compares several sources at fixed $\lambda_Q=1$, dividing by $NK$ to keep the curves visible on the same axes. Their one-loop colour factor is proportional to $N$.

\begin{figure}[htbp]
 \centering
 \includegraphics[width=.9\textwidth]{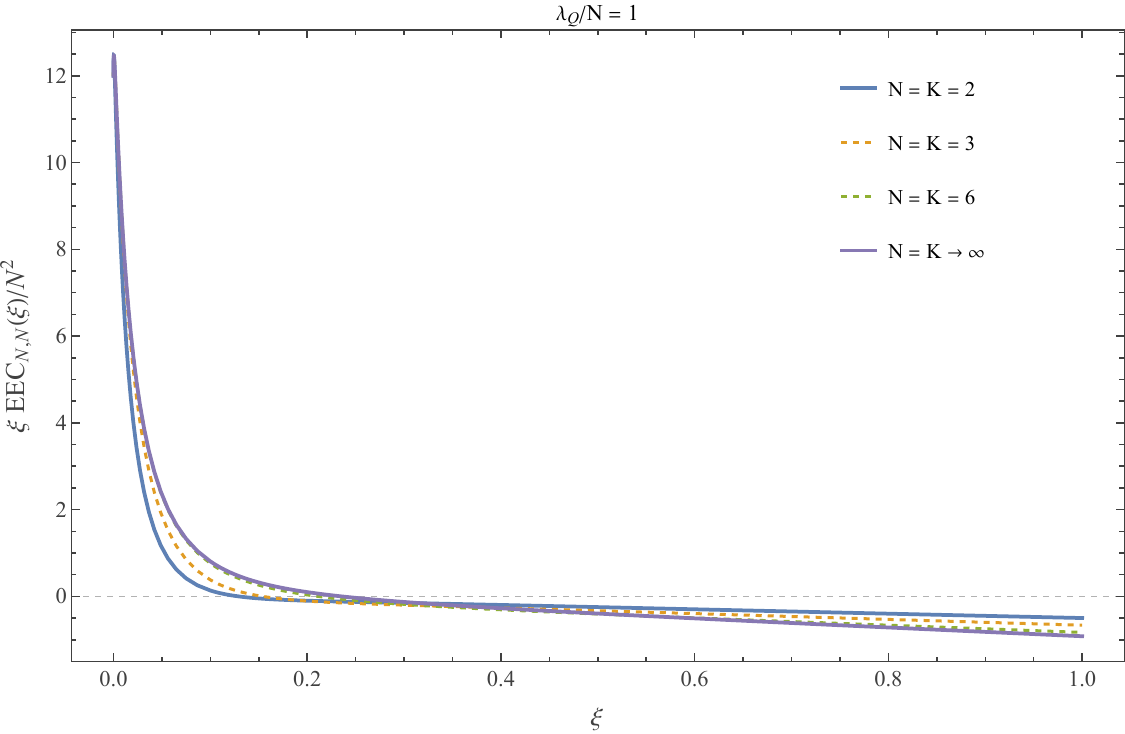}
 \caption{The equal-rank sequence $K=N$ and its infinite-rank limit
 at fixed $\lambda_Q/K=1$, showing $\xi\EEC_{N,N}(\xi)/N^2$.}
 \label{fig:eec-continuum}
\end{figure}

For the equal-rank sequence $K=N\to\infty$, we take the same
large-charge limit as~\cite[section~5.4.2]{Brown:2025cbz}, holding
$m g_{\rm YM}^2$ fixed. Since $\lambda_Q=mK g_{\rm YM}^2/(16\pi^2)$,
$\lambda_Q/N=m g_{\rm YM}^2/(16\pi^2)$ is then fixed automatically. The masses
$M_s^2=4\lambda_Q\sin^2(\pi s/N)/N$ then sample a continuous
distribution. Writing $\eec(\xi;M^2)$ for the single-mode function,
the sum becomes
\begin{equation}
 \lim_{\substack{N\to\infty\\\lambda_Q/N=1}}
 \frac{\EEC_{N,N}(\xi;\lambda_Q)}{N^2}
 =\int_0^1d\rho\,
 \eec\bigl(\xi;4\sin^2(\pi\rho)\bigr).
 \label{eq:finite-coupling-continuum}
\end{equation}
Figure~\ref{fig:eec-continuum} shows the finite-rank curves approaching this limit. The positive collinear region and negative back-to-back region persist, while the zero moves with the mass distribution. Holding $\lambda_Q$ itself fixed would instead make all masses vanish, and $\EEC_{N,N}/N$ would then reduce to the one-loop result. 

\begin{figure}[htbp]
 \centering
 \includegraphics[width=.9\textwidth]{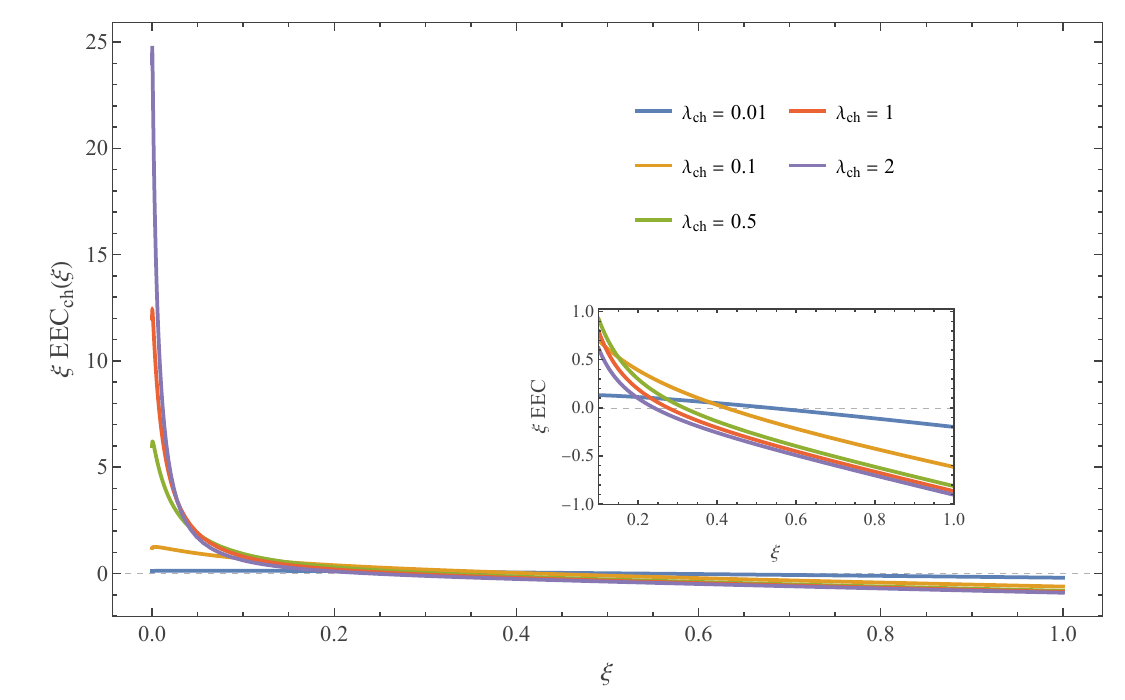}
 \caption{The coherent EEC showing $\xi\EEC_{\rm{ch}}(\xi)$ for different values of $\lambda_{\rm{ch}}$ as defined in \eqref{eq:coherent-coupling}.}
 \label{fig:eec-continuum-2}
\end{figure}

Next in figure \ref{fig:eec-continuum-2} we consider another averaging over all the massive modes in the form of the coherent correlator of section \ref{sec:coherent-states}. This is related to $\EEC_{2,2}$ by an integral transform of the coupling\footnote{In principle we should perform it before taking the double discontinuity and perform the integration over $w,\bar{w}$ but the two operations commute.}
\begin{equation}
\mathrm{EEC}_{\rm ch}(\xi;\lambda_{\rm ch})
=
\int_0^1 dr\,w(r)\,
\mathrm{EEC}_{2,2}\!\left(\xi;8\lambda_{\rm ch}r^2\right),
\end{equation}
with $w(r)$ given in  \eqref{eq:weight}, 
and the coupling $\lambda_{\rm{ch}}$ is define as \eqref{eq:coherent-coupling}. 
In this case, similarly to the large $K=N$ case, we see that the correlator smoothly interpolates between the collinear and back-to-back region. We will further discuss this in section~\ref{sec:canonical-strong}. 

Finally, to check our results we verified that the sum rules
\begin{equation}
\int_0^1\!d\xi\,\EEC(\xi;\lambda)=0,
\qquad
\int_0^1 \!d\xi\,\xi\EEC(\xi;\lambda)=0.
\label{eq:sum_rules_endp}
\end{equation}
vanish numerically at finite coupling. As in the one-loop case \eqref{eq:ps-one-loop-sum-rules}, the unweighted integral is understood distributionally, with the collinear $1/\xi$ singularity treated using a plus prescription and the endpoint contact term included. 

\subsection{Strong coupling and small-angle scaling}
\label{sec:canonical-strong}
We now consider the strong coupling expansion of the EEC. 
We start by studying the $N=K=2$ case at a fixed nonzero detector angle. In order to do so we need to expand the integral representation \eqref{eq:ps-I} in the only region that contributes in this regime which is the small region around  $w=0$ and $\bar{w}=1$. We find that the expansion takes the following form, 
\begin{equation}
 \EEC_{2,2}(\xi;\lambda_Q)
 =-2+\frac1{12\pi^2\lambda_Q\xi^3}
 +\frac{\mathcal B_2(\xi)}{\lambda_Q^2}
 +O\!\left(\frac{\log\lambda_Q}{\lambda_Q^3}\right),
 \label{eq:results-strong-generic}
\end{equation}
where
\begin{equation}
 \mathcal B_2(\xi)=
 \left(\frac1{240\pi^2}+\frac1{40\pi^4}\right)\frac{2\xi-3}{\xi^4}
 -\frac3{10\pi^4}
 \frac{(1+4\xi+\xi^2)\log\xi+3(1-\xi^2)}{(1-\xi)^5}.
 \label{eq:results-strong-second}
\end{equation}
Its apparent singularity at $\xi=1$ is removable. It is important to note that the expansion
breaks down towards the collinear endpoint as the constant term dominates only when $\lambda_Q\xi^3\gg1$.

\begin{figure}[htbp]
 \centering
 \includegraphics[width=.98\textwidth]{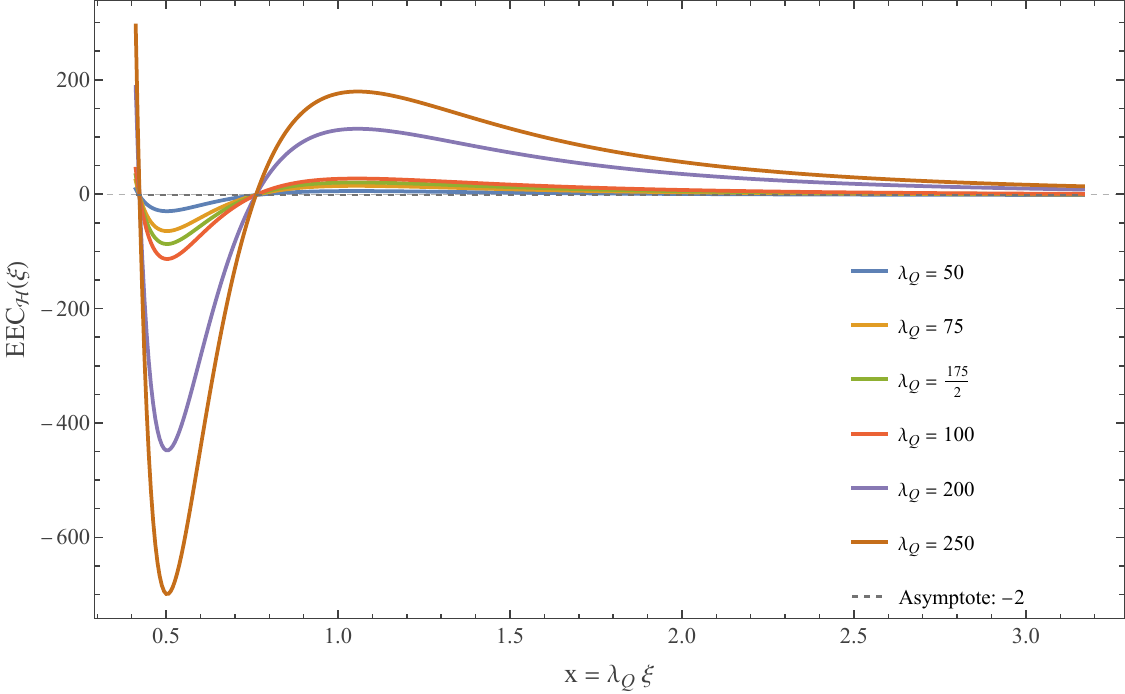}
 \caption{The $N=K=2$ EEC against ${\bf x}=\lambda_Q\, \xi$
 for $\lambda_Q\in\{50,75,87.5,100,200,250\}$. The dashed line marks the strong-coupling
 plateau $-2$.}
 \label{fig:eec-strong-scaling}
\end{figure}

\begin{figure}[htbp]
 \centering
 \includegraphics[width=.98\textwidth]{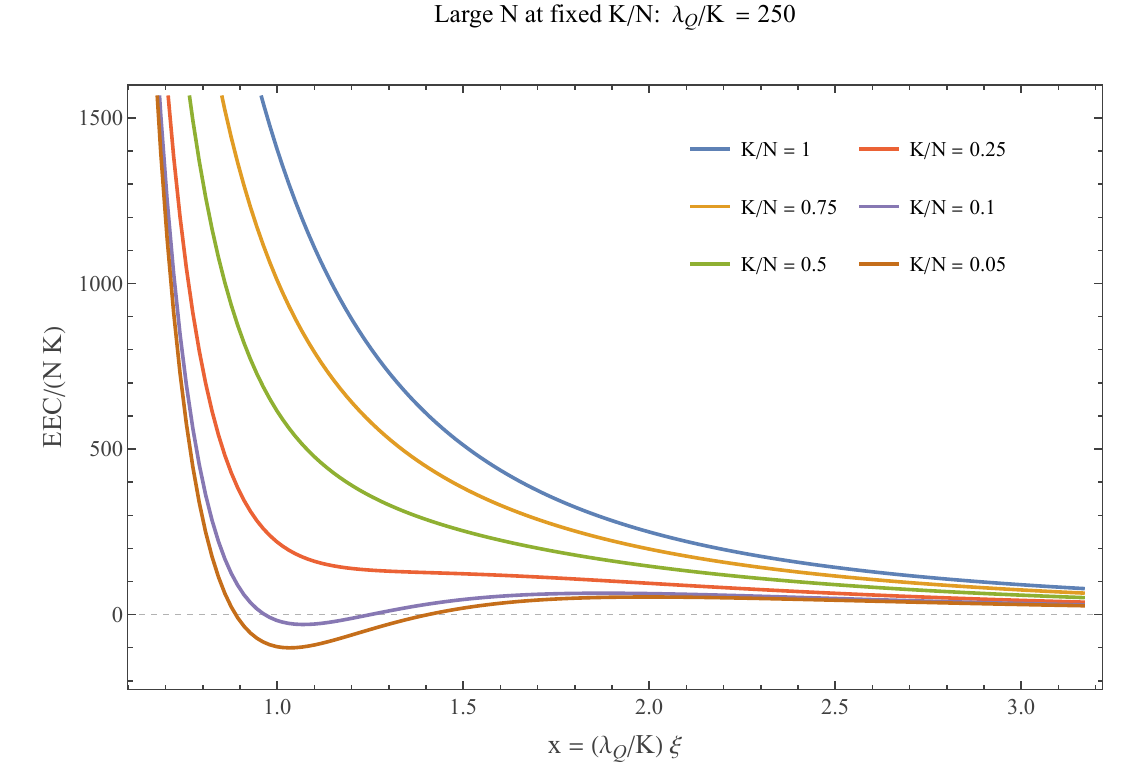}
 \caption{EEC in the large $N$ limit, with a fixed $K/N$.}
 \label{fig:eec-largeN-fixedK/N}
\end{figure}

To resolve the shrinking angular region, we introduce an effective coupling ${\bf x}$, defined as
\begin{equation}
{\bf x} =\lambda_Q\, \xi \, ,
\end{equation}
and we will consider a strong coupling expansion of the EEC with the effective coupling ${\bf x}$ fixed 
\begin{equation}
 \EEC_{2,2}({\bf x}/\lambda_Q;\lambda_Q)
 =\lambda_Q^2\Psi({\bf x})+ O(\lambda_Q^1) \, . 
 \label{eq:results-double-scaling}
\end{equation}
As we see in figure \ref{fig:eec-strong-scaling}, the $\EEC$ has an initial dip whose amplitude scales as $\lambda_Q^2$ at fixed ${\bf x}$. In particular, around $\bf x = 0$ the profile is described by $\Psi({\bf x})\sim24/{\bf x}$. The EEC then rebounds to positive values when $2{\bf x}$ is of order one, which is followed by an asymptotic approach to the strong coupling plateau as ${\bf x}$ continues to grow. The large $\bf x$ tail scales as $\Psi({\bf x})\sim1/(12\pi^2 {\bf x}^3)$, which we can see from the strong coupling results of \eqref{eq:results-strong-generic}, reaching the plateau when ${\bf x} \gg \lambda_Q^{2/3}$, or equivalently $\xi \gg \lambda_Q^{-1/3}$.
A similar angular crossover occurs in QCD jet correlators, where varying the angular resolution probes the transition between partonic and hadronic dynamics~\cite{Komiske:2022enw,Chicherin:2023gxt}. Here, the scale follows from the massive
propagator, whose collinear region keeps $M_1^2\xi$ finite.
Since $\xi\sim \theta^2/4$ for small $\theta$, this probes opening angles of order
$\lambda_Q^{-1/2}$.
Interestingly we see that this behaviour is not uniform in $N,K$. The onset of the negative dip in figure~\ref{fig:eec-largeN-fixedK/N} as the ratio $K/N$ decreases can be thought of as a competition between the two mass terms in \eqref{eq:finite-coupling-mode-sum}. The sum over mass modes $s \in [1, K-1]$ in the limit of large $N$ averages the EEC to give the smooth shape observed in the $N=K$ case. As $N$ becomes increasingly large compared to $K$, the second mass term in the sum dominates for which the masses are degenerate, which leads to an enhancement of the profile seen in figure~\ref{fig:eec-strong-scaling} and the emergence of the negative dip. Analogously also $\EEC_{\rm{ch}}$ has a similar smooth interpolating behaviour as the large $N=K$ scaling as both average over a mass distribution.   

\subsection{Endpoint limits}
\label{sec:eec-endpoint-limits}
We now want to focus on the collinear and back-to-back regions of the $\EEC$. We start by studying the collinear region at weak coupling. For collinear detectors with $\xi\to0^+$, the one-loop angular coefficient behaves as
\begin{equation}
 \eec^{(1)}(\xi)=\frac{12}{\xi}-24\log\xi-26
 +O(\xi\log\xi),\qquad \xi\to0^+.
 \label{eq:results-collinear-one-loop}
\end{equation}
In order to see if this has any higher order correction we also expand the two-loop scalar function
\eqref{eq:results-two-loop} and by applying $\mathbb D_\xi$ we obtain
\begin{equation}
\begin{aligned}
 \eec^{(2)}(\xi)={}&
 48(3-\zeta_2)\log^2\xi
 +(192\zeta_3+40\zeta_2-72)\log\xi\\
 &+460-120\zeta_2-176\zeta_3-216\zeta_4
 +O(\xi\log^2\xi).
\end{aligned}
 \label{eq:results-collinear-two-loop}
\end{equation}
The two-loop coefficient therefore grows only as $\log^2\xi$. Higher loop orders are also subleading in this limit. The detector OPE explains this one-loop exactness~\cite{Chicherin:2023gxt,Brown:2024yvt}. At fixed $\lambda_Q$, the $L$-loop term in the anomalous dimension of the exchanged spin-three light-ray operator on the leading-twist trajectory is suppressed by $\Delta_{\mathcal H}^{-L}$, whereas its tree-level heavy-heavy matrix element grows as $\Delta_{\mathcal H}$. Only the first anomalous dimension term then leaves an unsuppressed logarithmic singularity in the large-charge correlator. Higher-loop matrix-element corrections are finite in this OPE limit. The detector transform then gives the following behaviour
\begin{equation}
 \lim_{\xi\to0^+}\xi\eec^{(L)}(\xi)=12\delta_{L,1},
 \qquad
 \lim_{\xi\to0^+}\xi\EEC_{\mathcal H}(\xi;\lambda_Q)=12N\lambda_Q.
 \label{eq:results-collinear-endpoint}
\end{equation}
Our numerical results in figure~\ref{fig:eec-finite-coupling} support this result at finite coupling. For $N=K=2$ it fixes the $24/{\bf x}$ behaviour in~\eqref{eq:results-double-scaling}, and hence the $\lambda_Q^2$ growth seen in the scaled plot.

At the regular back-to-back $\xi \to 1^-$ endpoint the perturbative coefficients are finite. They can be computed directly from the definition \eqref{eq:ps-I} by setting $\xi=1$. Through four loops we have
\begin{equation}
 \EEC_{\mathcal H}(1;\lambda_Q)=\sum_{L=1}^4d_{\mathcal H,L}\lambda_Q^L\, e_L+O(\lambda_Q^5),
 \label{eq:results-weak-back-to-back}
\end{equation}
with
\begin{align}
 e_1={}&-14-8\zeta_2,\notag\\
 e_2={}&\frac{55}{3}+\frac{644}{5}\zeta_2
       -16\zeta_4-8\zeta_3,\notag\\
 e_3={}&-\frac{1277}{216}-\frac{16381}{54}\zeta_2
       +\frac{1807}{2}\zeta_4+57\zeta_6-\frac{1289}{9}\zeta_3\notag\\
      &-804\zeta_2\zeta_3+32\zeta_3^2+30\zeta_5,\notag\\
e_4={}&-\frac{107227}{93312}
+\frac{8278501}{23328}\zeta_2
-\frac{7308403}{864}\zeta_4
+\frac{349753}{432}\zeta_6
-\frac{12022}{45}\zeta_8
+\frac{184}{5}\zeta_{3,5}
\notag\\
&-\frac{1191931}{3888}\zeta_3
+\frac{382985}{108}\zeta_2\zeta_3
-275\zeta_4\zeta_3
+\frac{404}{9}\zeta_3^2
+56\zeta_2\zeta_3^2
\notag\\
&+\frac{455873}{216}\zeta_5
+\frac{5972}{3}\zeta_2\zeta_5
-296\zeta_3\zeta_5
+\frac{23681}{12}\zeta_7\, .
 \label{eq:results-back-to-back-coefficients-three}
\end{align}

In the strong coupling expansion, the back-to-back endpoint approaches the same constant as the generic-angle result in \eqref{eq:results-strong-generic}. Again using $N=K=2$, the asymptotic expansion reads
\begin{equation}
 \EEC_{2,2}(1;\lambda_Q)=-2+\frac1{12\pi^2\lambda_Q}
 -\left(\frac1{240\pi^2}+\frac3{200\pi^4}\right)\frac1{\lambda_Q^2}
 +O(\lambda_Q^{-3}\log\lambda_Q).
 \label{eq:sc-first-terms}
\end{equation}
The next term is $(A_3\log\lambda_Q+B_3)/\lambda_Q^3$, with
\begin{align}
 A_3&=\frac{3(\pi^4+90)}{2800\pi^6},\\
 B_3&=2A_3(\gamma_{\rm E}+\log\pi)
 +\frac{9(\pi^4+60)}{2800\pi^6}\log2
 +\frac{27}{280\pi^6}S_2
 \nonumber\\
 &\quad-\frac{5471}{1176000\pi^2}
 -\frac{3}{560\pi^4}
 -\frac{6369}{19600\pi^6}\, ,
\end{align}
where $\gamma_E$ is the Euler--Mascheroni constant and $S_2=-\sum_{r=2}^{\infty}r^{-4}\log(1-r^{-2})$.

\begin{figure}[htbp]
 \centering
 \includegraphics[width=.8\textwidth]{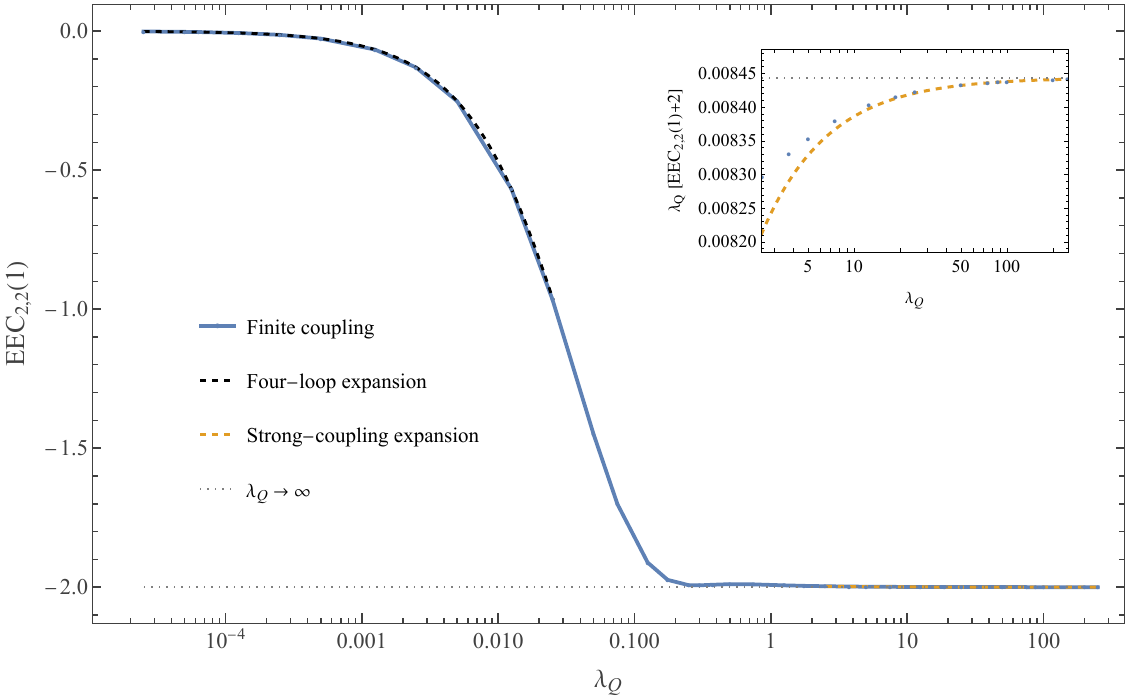}
 \caption{The regular $N=K=2$ back-to-back EEC as a function of
 $\lambda_Q$, compared with the four-loop weak-coupling result and
 the strong-coupling expansion~\eqref{eq:sc-first-terms}. The inset plots the correction to the leading $-2$ term at large $\lambda_Q$.}
 \label{fig:eec-finite-coupling-endpoint}
\end{figure}

The expansion is derived in appendix~\ref{app:strong-expansion}. In summary, we start by deforming the detector integrals from \eqref{eq:ps-I} onto compact contours $0 < \alpha, \beta < 2\pi$. The inverse-mass corrections arise from the \(n=0\) term in the worldline-instanton expansion of \eqref{eq:worldline-instanton-rep}, which ceases to be exponentially suppressed near the contour boundaries. Matching the boundary and interior expansions produces at most one logarithm at each order, starting at $\lambda_Q^{-3}$.
Figure~\ref{fig:eec-finite-coupling-endpoint} compares the numerical endpoint with the four-loop weak-coupling result and the strong-coupling expansion~\eqref{eq:sc-first-terms} showing perfect agreement.

\section{Conclusion and discussion}
\label{sec:con and out}

In this work, we studied EECs in heavy half-BPS states of $\mathcal N=4$ SYM in the large-charge ’t Hooft limit. The localization of the Fourier integral at large charge simplifies the detector transform, allowing us to express the dynamical contribution to the EEC in terms of a double discontinuity of the reduced four-point function. Combining this representation with the all-orders results for the corresponding HHLL correlators, we determine the EEC at finite effective coupling.

We applied this framework to obtain the angular dependence of the EEC explicitly through three loops. At each loop order, the dependence on the heavy source is contained in a colour factor, while the remaining angular function is universal within the class of correlators considered here. For canonical operators, the resummed correlator further determines the EEC at finite $\lambda_Q$ as a sum over the massive modes of the Coulomb-branch background. Combining analytic results with numerical evaluation, we studied its dependence on the effective coupling and the separation between the detectors at generic angles and towards the angular endpoints. We also determined the coherent-state colour factors from supersymmetric localization and expressed the corresponding correlator as an integral transform of the resummed ladder building block.

At the regular back-to-back endpoint, we derived a systematic strong-coupling expansion. The single-mode dynamical contribution approaches a coupling-independent constant with inverse-coupling corrections, and logarithms first appearing at order $\lambda_Q^{-3}$. These corrections arise from boundary regions of the detector integrals, where the exponential suppression of the Euclidean propagator is no longer uniform. This illustrates how the detector limit can retain contributions that are not captured by a strong-coupling expansion at fixed Euclidean cross ratios. We also studied the interplay between angular separation and coupling through the scaling variable ${\bf x}=\lambda_Q\xi$. This resolves the shrinking collinear region and connects its one-loop-exact singularity to the small-angle tail of the strong-coupling expansion, making explicit the noncommutativity of the two limits.

Several questions follow naturally from these results. An interesting direction is to understand the noncommutativity of the collinear limit and the strong-coupling expansion from the light-ray OPE. The leading collinear behaviour is one-loop exact at finite $\lambda_Q$, whereas the strong-coupling expansion at fixed angle does not reproduce this behaviour term by term. It would be useful to identify how the relevant light-ray OPE contributions reorganize at strong coupling and to develop a uniform description connecting the two regimes.

The framework developed here could be extended to a broader class of heavy states, higher-point energy correlators, and subleading orders in the inverse charge. A detailed comparison between canonical and coherent sources would help distinguish general features of large-charge energy flow from those sensitive to the choice of state. Another direction is to study the large-charge limit at fixed Yang--Mills coupling, beyond the weak-coupling double scaling considered here, and determine which features of the EEC persist in that regime. Existing results for integrated correlators~\cite{Paul:2023rka,Brown:2023why} could constrain this extension and its finite-charge corrections, helping to establish the range of validity of the leading semiclassical description and clarify its relation to other large-charge and large-$N$ limits.

\section*{Acknowledgements}
We thank Carlos Bercini, Murat Kolo\u{g}lu, Rodolfo Russo, Bo Wang, and Alexander Zhiboedov for helpful discussions. MJ is supported by a Science and Technology Facilities Council (STFC) studentship. CW and AG are supported by a Royal Society University Research Fellowship No.~URF$\backslash$R$\backslash$221015 and a STFC Consolidated Grant, ST$\backslash$X00063X$\backslash$1 ``Amplitudes, strings \& duality". AG is also supported by a Royal Society University Research Fellowship URF$\backslash$R1$\backslash$241371. AG also acknowledges support from OpenAI through the ChatGPT for Academic Researchers program.  

\appendix
\begingroup
\raggedbottom
\setlength{\jot}{3pt}

\section{Angular iterated integrals}
\label{app:angular-iterated-integrals}
In this appendix we present a more compact way to write the perturbative results for the $\EEC$. We find that it is possible to rationalise all the letters and write the expansion in terms of real iterated integrals with four kernels if we write the expansion directly in terms of the angle between the two detectors. 
Throughout we use the following definitions for the angles and functions.
\begin{equation}
 \xi=\sin^2\phi,\qquad 0<\phi<\frac\pi2,\qquad
 \mathcal J_w(\xi):=J_w(\arcsin\sqrt \xi)\, , 
 \label{eq:angular-variable}
\end{equation}
The functions $J$ are defined in the following recursive way
\begin{equation}
 J_\emptyset(\phi)=1,\qquad
 J_{aw}(\phi)=\int_0^\phi k_a(t)J_w(t)\,dt,
 \label{eq:angular-definition}
\end{equation}
with kernels
\begin{equation}
 k_C(\phi)=\cot\phi,\qquad k_T(\phi)=\tan\phi,\qquad
 k_E(\phi)=1,\qquad k_S(\phi)=\sec\phi.
 \label{eq:angular-kernels}
\end{equation}
As usual the leftmost letter is the outermost integration. For example,
$J_{TC}$ integrates $\tan t\,J_C(t)$ from $0$ to $\phi$.
In the case $J_{C^n}$ we need to implement a lower integration point prescription that gives $J_{C^n}=\log^n(\sin\phi)/n!$. Every other non-empty word vanishes at
$\phi=0$. As in the canonical case we can define the weight of a function by the word length. 

The weight-one functions are
\begin{equation}
\begin{aligned}
 \mathcal J_C(\xi)&=\tfrac12\log \xi,
 &\mathcal J_T(\xi)&=-\tfrac12\log(1-\xi),\\
 \mathcal J_E(\xi)&=\arcsin\sqrt \xi,
 &\mathcal J_S(\xi)&=\operatorname{arctanh}\sqrt \xi.
\end{aligned}
 \label{eq:angular-weight-one}
\end{equation}
In particular, $\mathcal J_S(\xi)=\tfrac12\log[(1+\sqrt \xi)/(1-\sqrt \xi)]$.
Writing $\omega_a=k_a(\phi)\,d\phi$, the kernels in the variable $\xi$ are
\begin{equation}
\begin{aligned}
 \omega_C&=\frac{d\xi}{2\xi},
 &\omega_T&=\frac{d\xi}{2(1-\xi)},\\
 \omega_E&=\frac{d\xi}{2\sqrt{\xi(1-\xi)}},
 &\omega_S&=\frac{d\xi}{2\sqrt \xi(1-\xi)}.
\end{aligned}
 \label{eq:angular-root-letters}
\end{equation}

As for harmonic polylogarithms (HPLs)~\cite{Remiddi:1999ew}, differentiation
removes the first letter, and products obey the shuffle rule:
\begin{equation}
 \frac{d}{d\phi}J_{aw}=k_aJ_w,\qquad
 J_uJ_v=\sum_{w\in\operatorname{Sh}(u,v)}J_w.
 \label{eq:angular-shuffle}
\end{equation}
A shuffle interleaves two words without changing the order within either
word; repeated words are counted with multiplicity. Thus
$\mathcal J_C^2=2\mathcal J_{CC}$ and
$\mathcal J_C\mathcal J_T=\mathcal J_{CT}+\mathcal J_{TC}$.
After expanding a product in this way, integration against $k_a$ simply
adds the letter $a$ at the front.

For words containing only $C,T$ we have a further simplification:
\begin{equation}
 \mathcal J_{a_1\cdots a_n}(\xi)
 =2^{-n}H_{\sigma(a_1)\cdots\sigma(a_n)}(\xi),\qquad
 \sigma(C)=0,\quad\sigma(T)=1,
 \label{eq:angular-hpl-subsector}
\end{equation}
where the HPL kernels are $d\xi/\xi$ and $d\xi/(1-\xi)$. For example,
$\mathcal J_{CT}(\xi)=\operatorname{Li}_2(\xi)/4$ and
$\mathcal J_{CCT}(\xi)=\operatorname{Li}_3(\xi)/8$.
We can rationalise all the four kernels at the same time in the half-angle variable
\begin{equation}
 t=\tan\frac\phi2=\frac{\sqrt \xi}{1+\sqrt{1-\xi}},\qquad 0<t<1:
 \label{eq:angular-cyclotomic-variable}
\end{equation}
\begin{equation}
\begin{aligned}
 \omega_C&=\frac{dt}{t}-\frac{2t\,dt}{1+t^2},
 &\omega_T&=\frac{2t\,dt}{1-t^2}+\frac{2t\,dt}{1+t^2},\\
 \omega_E&=\frac{2\,dt}{1+t^2},
 &\omega_S&=\frac{2\,dt}{1-t^2}.
\end{aligned}
 \label{eq:angular-rational-alphabet}
\end{equation}
The denominators involve the cyclotomic polynomials
$\Phi_1(t)=t-1$, $\Phi_2(t)=t+1$ and $\Phi_4(t)=1+t^2$~\cite{Ablinger:2011te}.
The poles are $0,\pm1,\pm i$, so the $J_w$ are real combinations of
level-four cyclotomic polylogarithms. As we can see $C,T$ describe the harmonic polylogarithms letters $\{0,1\}$ while the $E,S$ describe the complex and square root letters. As we will see this rewriting is useful in reducing the size of the expressions. The normalization at
$\phi=0$ must be kept when changing variables; for example,
\begin{equation}
 J_C=\log t-\log(1+t^2)+\log2.
 \label{eq:angular-basepoint-conversion}
\end{equation}
The $E$ letter also gives Clausen functions. With
$\operatorname{Cl}_2(\theta)=\operatorname{Im}\operatorname{Li}_2(e^{i\theta})$,
\begin{equation}
 J_{EC}(\phi)=-\tfrac12\operatorname{Cl}_2(2\phi)-\phi\log2.
 \label{eq:angular-clausen-example}
\end{equation}
Further integrations with $E,C,T$ give generalized Clausen and log-sine
functions~\cite{Borwein:2011mq,Umezawa:2019xyv}.

\subsection{Endpoint expansions}
\label{app:angular-endpoints}
As it will be useful for computing the endpoint behaviour let us briefly discuss the $\xi->0$ and $\xi->1$ expansion for $\mathcal{J}$.
The endpoint series follow from the same defining integrals. In the
variable $\xi$, their recursion is
\begin{equation}
 \frac{d\mathcal J_{aw}(\xi)}{d\xi}=f_a(\xi)\mathcal J_w(\xi),
 \qquad f_a(\xi)\,d\xi=\omega_a,
 \label{eq:angular-z-recursion}
\end{equation}
with $\omega_a$ given in~\eqref{eq:angular-root-letters}.
Near $\xi=0$, expand the kernel and the shorter word, then integrate each
term. The lower-limit prescription fixes the integration constants.
The result has the form
\begin{equation}
 \mathcal J_w(\xi)=\sum_{n\geq0}\sum_{k=0}^{|w|}
 c^{(0)}_{w,n,k}\,\xi^{n/2}\log^k \xi.
 \label{eq:angular-zero-series}
\end{equation}
Half-integer powers come from $E,S$; words made only of $C,T$ have
integer powers. For example,
\begin{equation}
\begin{gathered}
 \mathcal J_E(\xi)=\sqrt \xi+\tfrac16\xi^{3/2}+O(\xi^{5/2}),\qquad
 \mathcal J_S(\xi)=\sqrt \xi+\tfrac13\xi^{3/2}+O(\xi^{5/2}),\\
 \mathcal J_C(\xi)=\tfrac12\log \xi,\qquad
 \mathcal J_T(\xi)=\tfrac \xi2+\tfrac{\xi^2}4+O(\xi^3),\\
 \mathcal J_{EC}(\xi)=\sqrt \xi\left(\tfrac12\log \xi-1\right)
 +\xi^{3/2}\left(\tfrac1{12}\log \xi-\tfrac1{18}\right)
 +O(\xi^{5/2}\log \xi).
\end{gathered}
 \label{eq:angular-zero-examples}
\end{equation}

Near $\xi=1$, set $y=1-\xi$ and use
\begin{equation}
 \frac{d\mathcal J_{aw}(1-y)}{dy}
 =-f_a(1-y)\mathcal J_w(1-y).
 \label{eq:angular-one-recursion}
\end{equation}
Expanding and integrating in $y$ gives the series up to a constant for
each word. These constants are fixed by the original integrals:
\begin{equation}
 K_w=\operatorname{Reg}_{y\to0}\mathcal J_w(1-y),
 \qquad K_\emptyset=1.
 \label{eq:angular-connection-constants}
\end{equation}
Here $\operatorname{Reg}$ extracts the constant term after subtracting the
divergent logarithms. For example, $K_E=\pi/2$, $K_S=\log2$ and
$K_{CT}=\zeta_2/4$. They can also be found by matching the series to the
integrals at an interior point $0<\xi<1$. This method works for all four letters.
The first few expansions are
\begin{equation}
\begin{aligned}
 \mathcal J_E(1-y)&=\tfrac\pi2-\sqrt y
 -\tfrac16y^{3/2}+O(y^{5/2}),\\
 \mathcal J_C(1-y)&=-\tfrac12y-\tfrac14y^2+O(y^3),\\
 \mathcal J_T(1-y)&=-\tfrac12\log y,\\
 \mathcal J_S(1-y)&=\log2-\tfrac12\log y-\tfrac14y
 -\tfrac3{32}y^2+O(y^3),\\
 \mathcal J_{EC}(1-y)&=-\tfrac\pi2\log2
 +\tfrac16y^{3/2}+O(y^{5/2}),\\
 \mathcal J_{CT}(1-y)&=\tfrac{\zeta_2}{4}
 +\tfrac14y(\log y-1)+O(y^2\log y).
\end{aligned}
\label{eq:angular-one-examples}
\end{equation}
In particular, words such as $T$ and $S$ diverge logarithmically;
$K_w$ is their constant term, rather than a finite endpoint value.

\subsection{One- and two-loop EEC from $\langle \mathcal O_2 \mathcal O_2\mathcal O_2\mathcal O_2 \rangle$}
\label{app:2222-two-loop}

The EEC sourced by operator $\mathcal O_2$ follows from the
$\langle \mathcal O_2 \mathcal O_2\mathcal O_2\mathcal O_2 \rangle$ correlator by superconformal symmetry. We show that the
one- and two-loop result of \cite{Belitsky:2013ofa} can be rewritten in a relatively compact form using the functions we introduced above.  Using their normalization, we have
\begin{equation}
\begin{aligned}
 F(\xi;\lambda_Q)&:=4\xi^2(1-\xi)\EEC_{2222}(\xi;\lambda_Q)\\
 &=2N_c\lambda_QF^{(1)}(\xi)
 +\left(2N_c\lambda_Q\right)^2F^{(2)}(\xi)+O(\lambda_Q^3)\, .
\end{aligned}
 \label{eq:2222-eec-normalization}
\end{equation}
Here $\Delta_{\mathcal O_2}=2$, so that $\lambda_Q=2g_{\rm YM}^2/(16\pi^2)$,
and $\chi=2\theta$ is the detector angle. These are the ordinary functions
on $0<\xi<1$, with endpoint contact terms omitted. The normalization in
the main text is twice this angular distribution:
\begin{equation}
 \EEC_2(\xi;\lambda_Q)=2\EEC_{2222}(\xi;\lambda_Q)
 =\frac{F(\xi;\lambda_Q)}{2\xi^2(1-\xi)}.
 \label{eq:2222-eec-main-text-normalization}
\end{equation}

Equations~(13)--(15) of \cite{Belitsky:2013ofa} become
\begin{equation}
 F^{(1)}(\xi)=2\mathcal J_T(\xi),\qquad
 F^{(2)}(\xi)=(1-\xi)\mathcal W_2(\xi)+\mathcal W_3(\xi),
 \label{eq:2222-angular-result}
\end{equation}
where $\mathcal W_2$ and $\mathcal W_3$ have weights two and three.
All angular functions below have argument $\xi$:
\begin{align}
 \mathcal W_2(\xi)={}&8\sqrt \xi\,\mathcal J_{SC}
 +8\xi\mathcal J_{CT}-8\mathcal J_{TC}
 -8(1-\xi)\mathcal J_{TT}+2\xi\zeta_2,
 \label{eq:2222-angular-weight-two}\\
 \mathcal W_3(\xi)={}&-\tfrac12\xi(1+4\xi)\zeta_3
 +2\xi^2\zeta_2\mathcal J_C
 +(2\xi^2+\xi-2)\zeta_2\mathcal J_T\notag\\
 &+8\xi(\xi-2)\mathcal J_{CCT}
 +4\xi(3-4\xi)(\mathcal J_{CTC}+\mathcal J_{TTC})\notag\\
 &-4\xi(2\xi+1)\mathcal J_{CTT}
 +4(2\xi^2-3\xi-4)\mathcal J_{TCT}
 -8(\xi^2+2)\mathcal J_{TTT}\notag\\
 &-4(1-\xi)(1+2\xi)\mathcal J_{SSC}.
 \label{eq:2222-angular-weight-three}
\end{align}
Only $\mathcal J_{SC}$ and $\mathcal J_{SSC}$ require the $S$ letter;
the remaining words use $C,T$.

To obtain this form, set $r=\sqrt \xi$ and
$L_r=\log[(1+r)/(1-r)]=2\mathcal J_S(\xi)$. The root-dependent
polylogarithms become
\begin{equation}
 \operatorname{Li}_2(\pm r)=\mathcal J_{CT}\pm\mathcal J_{CS}\, .
 \label{eq:2222-root-dilogarithms}
\end{equation}
Shuffle reduction then gives
\begin{equation}
 \operatorname{Li}_2(-r)-\operatorname{Li}_2(r)
 +\frac{\log \xi}{2}L_r
 =2\mathcal J_{SC}\, ,
 \label{eq:2222-root-weight-two}
\end{equation}
and
\begin{equation}
\begin{aligned}
 &L_r^2\log\frac{1-\xi}{\xi}
 -8\operatorname{Li}_3\!\bigl(\tfrac{r}{r-1}\bigr)
 -8\operatorname{Li}_3\!\bigl(\tfrac{r}{r+1}\bigr)\\
 &\qquad=16(\mathcal J_{CCT}+\mathcal J_{CTT}
 +\mathcal J_{TCT}+\mathcal J_{TTT}-\mathcal J_{SSC})\, .
\end{aligned}
 \label{eq:2222-root-weight-three}
\end{equation}
The normalization at $\xi=0$ fixes the constants in these identities.
The endpoint expansions, with $y=1-\xi$, agree with Eqs.~(16) and~(18)
of \cite{Belitsky:2013ofa}:
\begin{align}
 F^{(2)}(\xi)&=\xi\left[\log \xi-\tfrac12\zeta_3+\zeta_2-3\right]
 +O(\xi^2\log \xi),\qquad \xi\to0,\notag\\
 F^{(2)}(1-y)&=\tfrac12\log^3y+\frac{\pi^2}{4}\log y
 +\tfrac12\zeta_3+O(y\log^3y),\qquad y\to0.
 \label{eq:2222-angular-endpoints}
\end{align}

\subsection{Two-loop EEC at large charge in the planar limit}
\label{app:large-charge-two-loop}

We next consider the EEC of~\cite{Chicherin:2023gxt} related to the correlator $\langle \mathcal{O}_p \mathcal{O}_p \mathcal{O}_2 \mathcal{O}_2 \rangle$ in the planar limit. The author consider the large-charge limit in the following way: first take $N\to\infty$
at fixed $N \lambda_Q/p = N g_{\rm YM}^2/(16\pi^2)$ (i.e. the usual 't Hooft coupling), where $p$ is the dimension of $\mathcal{O}_p$, then take the source charge
$p\to\infty$. The detector angle $0<\xi<1$ is kept fixed. Their two-loop coefficient scales as
\begin{equation}
 \mathcal F_p(\xi;\lambda_Q)=\sum_{\ell\geq0}\left(\frac{4N \lambda_Q}{p}\right)^\ell\mathcal F_p^{(\ell)}(\xi)\, ,
 \qquad
 \mathcal F_p^{(2)}(\xi)
 =\frac{\varphi^{(2)}(\xi)}{p}+O(p^{-2})\, .
 \label{eq:large-charge-normalization}
\end{equation}
The result is given in their ancillary file
\texttt{asymptotics-EEC-largeK.m} and discussed in Sec.~5.4.
It includes the detector derivatives; endpoint contact terms are omitted.

We now compare its correlator with the HHLL correlator considered in this paper. Using the cross ratios of~\eqref{eq:uv-def},
\begin{equation}
 u=\frac{x_{12}^2x_{34}^2}{x_{13}^2x_{24}^2},\qquad
 v=\frac{x_{14}^2x_{23}^2}{x_{13}^2x_{24}^2} \, , 
\end{equation}
with integration measure $d^4x/(4\pi^2)$, their planar two-loop
expression~\cite[eqs.~(B.9)--(B.14)]{Chicherin:2023gxt} becomes
\begin{equation}
\begin{aligned}
 16x_{13}^2x_{24}^2F_{p>2,\mathrm{planar}}^{(2)}(x_i)
 ={}&2P^{(2)}(1/u,v/u)+\frac2uP^{(2)}(u,v)
     +\frac1vP^{(2)}(v,u)\\
 &+\frac{1+u}{2u^2}[P^{(1)}(u,v)]^2.
\end{aligned}
 \label{eq:results-planar-two-loop-input}
\end{equation}
The ladder integrals $P^{(2)}$ is defined in~\eqref{eq:ladder def}.
We see that this includes the crossed
two-loop ladders, rather than the combination
$(P^{(1)})^2+2P^{(2)}$ appearing in~\eqref{eq:hhll-master}.

With this set up, we  find the EEC in this particular limit of~\cite{Chicherin:2023gxt} takes the following form, 
\begin{equation}
\begin{aligned}
 \xi d^4\varphi^{(2)}(\xi)={}&
 b_\emptyset+b_C\mathcal J_C+b_{CC}\mathcal J_{CC}
 +b_{TC}\mathcal J_{TC}\\
 &+b_{CTC}\mathcal J_{CTC}+b_{TCC}\mathcal J_{TCC}
 +b_{CCTC}\mathcal J_{CCTC}+b_{CTCC}\mathcal J_{CTCC}\\
 &+b_{TCTC}\bigl(\mathcal J_{TCTC}
 +\tfrac{\zeta_3}{4}\mathcal J_T\bigr)\\
 &+b_{ECC}\bigl(\mathcal J_{ECC}+\mathcal J_{ETC}
 -\tfrac{\pi^2}{24}\mathcal J_E\bigr)\\
 &+b_{EECC}\bigl(\mathcal J_{EECC}+\mathcal J_{EETC}
 -\tfrac{\pi^2}{24}\mathcal J_{EE}\bigr)\, ,
\end{aligned}
\label{eq:large-charge-angular-result}
\end{equation}
where $d=1-\xi$. 
This expression contains 16 words, including $\mathcal J_\emptyset=1$.
The coefficient functions are
\begin{equation}
\begin{aligned}
 b_\emptyset={}&\tfrac18d^3(-3d^2+4d+1)
 +\tfrac{\pi^2}{96}d^2(78d^3-32d^2+14d-27)\\
 &-\tfrac{\pi^4}{720}\xi(55d^4-109d+162)
 -\tfrac{\zeta_3}{2}(12d^5-6d^4-3d^3+62d^2-134d+72),\\
 b_C={}&\tfrac14d^2(-18d^3+20d^2+2d+5)
 -\tfrac{\pi^2}{6}\xi(9d^4+2d^3+d^2+31d-36)\\
 &+6\xi(d^4-3d+6)\zeta_3,\\
 b_{CC}={}&\tfrac14d^2(52d^3-44d^2-4d+11)
 -\tfrac{2\pi^2}{3}\xi(d^4-4d+9),\\
 b_{TC}={}&\tfrac14d^2(130d^3-44d^2+10d+27)
 -\tfrac{\pi^2}{3}d\xi^2(d^2+d+1).
\end{aligned}
\label{eq:large-charge-coefficients-low}
\end{equation}
The remaining coefficients are
\begin{align}
 b_{CTC}={}&60 d^{5} - 44 d^{4} - 8 d^{3} - 96 d^{2} + 268 d - 144,\notag\\
 b_{TCC}={}&4 d \left(- 6 d^{4} + 4 d^{3} + d^{2} + 28 d - 36\right),\notag\\
 b_{CCTC}={}&16 \xi \left(- 2 d^{4} - d + 9\right),\notag\\
 b_{CTCC}={}&16 \xi \left(d^{4} + 5 d - 18\right),\notag\\
 b_{TCTC}={}&8 d \xi \left(d^{3} + 1\right),\notag\\
 b_{ECC}={}&\tfrac34\sqrt{d/\xi}\,(122\xi^2+75\xi-5),\notag\\
 b_{EECC}={}&48 \xi \left(\xi + 2\right).\label{eq:large-charge-coefficients-high}
\end{align}
The only explicit square root is in $b_{ECC}$. The $E$ words replace
the polylogarithms with complex square-root arguments.

The conversion uses $x=i\tan\phi$. For $q=x/(1+x)$,
\begin{equation}
 \operatorname{Li}_1(q)=\mathcal J_T(\xi)+i\mathcal J_E(\xi),
 \qquad
 d\log q=(\cot\phi-i)\,d\phi.
 \label{eq:large-charge-conversion}
\end{equation}
Thus higher polylogarithms are obtained by adding $C-iE$ at the front
of each word. The shuffle rule handles products, and complex-conjugate
terms combine into real functions.

From our expression, it is straightforward to consider the endpoint expansions, 
With $y=1-\xi$, we have, 
\begin{align}
 \varphi^{(2)}(\xi)
 &=
 \frac{3\log \xi-\frac32\zeta_3+3\zeta_2-\frac72}{\xi}
 +O(\log^2\xi),
 \qquad \xi\to0,
\\
 \varphi^{(2)}(1-y)
 &=
 \frac74\zeta_4+\frac{85}{6}\zeta_3
 +\frac{181}{180}\zeta_2+\frac{49}{144}+O(y),
 \qquad y\to0.
 \label{eq:large-charge-endpoints}
\end{align}
which completely agree with Eqs.~(5.24)--(5.25) in~\cite{Chicherin:2023gxt}.

It is worth noting that the expression in the ancillary file  \texttt{asymptotics-EEC-largeK.m} of~\cite{Chicherin:2023gxt}    contains 62 products of logarithms and
polylogarithms; the angular form we present here uses only 16 real words with a common argument.

\section{Three-loop ladder-product integral}
\label{app:three-loop-coefficients}

We give the coefficients of the three-loop scalar detector integral
in~\eqref{eq:results-three-loop}.
The function $Q(\theta)$ is defined in~\eqref{eq:results-angular-seed},
with $\phi=\theta/2=\arcsin\sqrt\xi$. Its successive primitives are
\begin{align}
 A(\xi)&=\mathcal J_{CEECC}+\mathcal J_{CEETC}
       +\mathcal J_{TEECC}+\mathcal J_{TEETC}
       -\frac{\zeta_2}{4}(\mathcal J_{CEE}+\mathcal J_{TEE}),\notag\\
 B(\xi)&=\mathcal J_{TCEECC}+\mathcal J_{TCEETC}
       +\mathcal J_{TTEECC}+\mathcal J_{TTEETC}
       -\frac{\zeta_2}{4}(\mathcal J_{TCEE}+\mathcal J_{TTEE}).
 \label{eq:results-angular-primitives}
\end{align}
Equivalently, in the angular variable $\phi$,
$dA/d\phi=(\cot\phi+\tan\phi)Q$ and $dB/d\phi=\tan\phi\,A$,
with $A(0)=B(0)=0$.

The full $C,T$ contribution is given below, with terms grouped by
word length from zero to six. All iterated integrals have argument $\xi$.
\begingroup
\allowdisplaybreaks[1]
\begin{align}
P_3(\xi)={}&
 \xi\Bigl[-8\zeta_3-38\zeta_5-\frac{\pi^4}{180}
 +\frac{\pi^6(2+101\xi)}{11340}
 +\frac{11\xi-5}{2}\zeta_3^2+3\pi^2\zeta_3\Bigr]
 \notag\\[4pt]
&-\xi\Bigl[6\xi\zeta_5+\pi^2\xi\zeta_3
 -\frac{\pi^4}{4}-\frac{4\pi^2}{3}+8\zeta_3\Bigr]\mathcal J_C
 \notag\\
&+\Bigl[(87\xi^2-124\xi+61)\zeta_5
 -\frac{\pi^2}{3}(16\xi^2-33\xi+17)\zeta_3
 \notag\\[-2pt]
&\hspace{12mm}
 +(24-8\xi)\zeta_3+\frac{\pi^4}{180}(45\xi-43)\Bigr]\mathcal J_T
 \notag\\[4pt]
&+4\pi^2\xi\mathcal J_{CC}
 -\xi\Bigl[\frac{\pi^4}{90}(65\xi-7)+120\zeta_3\Bigr]\mathcal J_{CT}
 \notag\\
&+\Bigl[-\frac{\pi^4}{90}(89\xi^2-90\xi+45)
 +\frac{4\pi^2}{3}(3\xi-5)+16\zeta_3\Bigr]\mathcal J_{TC}
 \notag\\
&-\Bigl[\frac{\pi^4}{90}(8\xi^2+15\xi-5)
 +8(15\xi-11)\zeta_3\Bigr]\mathcal J_{TT}
 \notag\\[4pt]
&+16\zeta_3\Bigl[
 \xi(\xi+2)\mathcal J_{CCT}
 +\xi(13\xi+1)\mathcal J_{CTT}
 \notag\\[-2pt]
&\hspace{12mm}
 +(14\xi^2-33\xi+17)\mathcal J_{TCT}
 +(\xi-1)(4\xi-5)\mathcal J_{TTT}\Bigr]
 \notag\\
&+4\xi(14\xi\zeta_3+5\pi^2-8)\mathcal J_{CTC}
 +8(6\xi^2\zeta_3+4\xi-\pi^2-4)\mathcal J_{TCC}
 \notag\\
&+\Bigl[-8(\xi^2-9\xi+4)\zeta_3
 +\frac{4\pi^2}{3}(15\xi-11)\Bigr]\mathcal J_{TTC}
 \notag\\[4pt]
&-\frac83\Bigl[
 3\xi(\pi^2\xi+4)\mathcal J_{CCTC}
 +6\pi^2\xi^2\mathcal J_{CTCC}
 +\pi^2\xi(13\xi+1)\mathcal J_{CTTC}
 \notag\\[-2pt]
&\hspace{12mm}
 +(\xi-3)(8\pi^2\xi-5\pi^2+12)\mathcal J_{TCTC}
 \notag\\[-2pt]
&\hspace{12mm}
 -2\pi^2(2\xi^2-6\xi+3)\mathcal J_{TTCC}
 +\pi^2(\xi-1)(4\xi-5)\mathcal J_{TTTC}\Bigr]
 \notag\\[4pt]
&+32\Bigl[
 4\xi\mathcal J_{CCCTC}
 +14\xi\mathcal J_{CCTCC}
 -15\xi\mathcal J_{CTCTC}
 \notag\\[-2pt]
&\hspace{12mm}
 +2(2\xi-1)\mathcal J_{TCCTC}
 +14(\xi-1)\mathcal J_{TCTCC}
 +(11-15\xi)\mathcal J_{TTCTC}\Bigr]
 \notag\\[4pt]
&+64\Bigl[
 \xi(\xi+2)\mathcal J_{CCTCTC}
 -\xi(\xi-2)\mathcal J_{CTCCTC}
 -14\xi^2\mathcal J_{CTCTCC}
 \notag\\[-2pt]
&\hspace{12mm}
 +\xi(13\xi+1)\mathcal J_{CTTCTC}
 -2(\xi^2-4\xi+2)\mathcal J_{TCCCTC}
 \notag\\[-2pt]
&\hspace{12mm}
 -14(\xi-1)^2\mathcal J_{TCCTCC}
 +(14\xi^2-33\xi+17)\mathcal J_{TCTCTC}
 \notag\\[-2pt]
&\hspace{12mm}
 +\xi(\xi+1)\mathcal J_{TTCCTC}
 -4\xi^2\mathcal J_{TTCTCC}
 +(\xi-1)(4\xi-5)\mathcal J_{TTTCTC}\Bigr].
 \label{eq:results-three-loop-CT}
\end{align}
\endgroup

These coefficients specify $\mathcal F^{(3)}$ in the normalization of~\eqref{eq:results-perturbative-expansion}. Applying $\mathbb D_\xi$ gives the three-loop energy-correlator coefficient.

\section{Back-to-back strong-coupling expansion}
\label{app:strong-expansion}

We derive the regular endpoint expansion for $N=K=2$, for which $M^2=2\lambda_Q$.
The large-charge limit is taken first, followed by $\xi\to1^-$ after acting with the detector derivatives, and finally we perform the large  $\lambda_Q$ expansion.
Writing $t_1\equiv t_1(1/V,U/V;\lambda_Q)$ with $\mu=2M_1$, define
\begin{equation}
\begin{gathered}
 \mathfrak D[F]
 =\lim_{\xi\to1^-}
 \frac{\mathbb D_\xi I[F](\xi)}{8\pi^2},
 \qquad
 f_1(\lambda_Q)
 :=\eec_{2,2;1}(1;\lambda_Q)
 =\mathfrak D[t_1^2-1],
 \\
 \EEC_{2,2}(1;\lambda_Q)=2f_1(\lambda_Q),
\end{gathered}
\end{equation}
where $I$ is the functional defined in \eqref{eq:ps-I}.

\subsection{Endpoint representation and contributing regions}
We deform the integration contours of the detector transform $\mathfrak D$ onto the compact parametrization
\begin{gather}
 w=\frac{1}{1-e^{i\alpha}},\qquad
 \bw=\frac{1}{1-e^{-i\beta}},\qquad
 U=w\bw,\qquad V=(1-w)(1-\bw),\nonumber\\
 \varphi=\frac{\alpha+\beta}{2},\qquad
 \delta=\frac{\alpha-\beta}{2},\qquad
 h=\frac{\cos\delta-\cos\varphi}{\sin\varphi},
 \qquad 0<\alpha,\beta<2\pi.
\end{gather}
The variables $U,V$ are the detector cross ratios from \eqref{eq:detector-uv}. Here $\varphi$ is a contour coordinate, distinct from the physical
detector opening angle $\theta$. The endpoint functional becomes
\begin{equation}
 \mathfrak D[F]=\frac1{8\pi^2}\operatorname{FP}
 \int_0^{2\pi}\!d\alpha\int_0^{2\pi}\!d\beta\,
 (\mathcal J^2-3\mathcal J+2)\widetilde F,
 \qquad \mathcal J=-h\partial_\varphi\big|_\delta
 =U\partial_U+V\partial_V.
 \label{eq:app-sc-contour}
\end{equation}
Here $\widetilde F$ means the function $F$ evaluated on the parametrized compact contours. The symmetric finite part $\operatorname{FP}$ retains the boundary-value prescription of the original discontinuity integral
\begin{equation}
 \operatorname{FP}x^{-k}
 =\frac12\bigl[(x+i0)^{-k}+(x-i0)^{-k}\bigr],
 \qquad k\geq1.
 \label{eq:app-sc-finite-part}
\end{equation}
Using primes to denote $\partial_\varphi$ at fixed $\delta$, we write
\begin{equation}
 (\mathcal J^2-3\mathcal J+2)\widetilde F
 =h^2\widetilde F''+(hh'+3h)\widetilde F'+2\widetilde F.
\end{equation}
Integration by parts then gives the weight
\begin{equation}
 \mathcal W=(h^2)''-(hh'+3h)'+2
 =h^2(1+3\cot^2\varphi),
\end{equation}
where we have used the fact that the boundary expression $h^2\widetilde F'+(3h-hh')\widetilde F$ vanishes on the outer edges for $\widetilde F=\widetilde t_1^{\,2}-1$. Using $\operatorname{FP}\int_0^{2\pi}d\alpha
\int_0^{2\pi}d\beta\,\mathcal W=8\pi^2$, we obtain
\begin{equation}
 f_1=-1+\frac1{8\pi^2}\operatorname{FP}
       \int_0^{2\pi}\!d\alpha\int_0^{2\pi}\!d\beta\,
       \mathcal W\widetilde t_1^{\,2}.
 \label{eq:app-sc-gf}
\end{equation}

Using the worldline instanton sum in~\eqref{eq:worldline-instanton-rep}, we analytically continue $\sigma$ to $i\delta$, which gives the representation
\begin{equation}
 \widetilde t_1
 =2h\mu\sum_{k\in\mathbb Z}
 \frac{\varphi+2\pi k}{R_k}K_1(\mu R_k),
 \qquad R_k^2=(\alpha+2\pi k)(\beta+2\pi k).
 \label{eq:app-sc-images}
\end{equation}
The integer $k$ labels images, while the subscript $1$ labels the
massive mode. For fixed positive $R_k$, the Bessel $K_1$ function decays exponentially. Only the terms $k=0,-1$ have arguments that vanish on the contour boundaries, and therefore only these terms contribute to the inverse powers of $\mu$:
\begin{equation}
\begin{gathered}
 \widetilde t_1\simeq\widetilde t_{1,0}+\widetilde t_{1,-1}
 =2h\mu\left[
 \frac{\varphi}{R_0}K_1(\mu R_0)
 +\frac{\varphi-2\pi}{R_{-1}}K_1(\mu R_{-1})\right],
\end{gathered}
\end{equation}
where $R_0^2=\alpha\beta$ and $R_{-1}^2=(2\pi-\alpha)(2\pi-\beta)$. These are the two contributions in the $n=0$ term of \eqref{eq:worldline-instanton-rep}, expressed here using the signed image index $k$. Squaring their sum gives
$\widetilde t_{1,0}^{\,2}+\widetilde t_{1,-1}^{\,2}
+2\widetilde t_{1,0}\widetilde t_{1,-1}$.
Reflection about the centre of the contour square exchanges the two
images and preserves $\mathcal W$. Their squared contributions to
$f_1$ are equal. The mixed term also contributes: $R_0$ and $R_{-1}$
vanish together at $(0,2\pi)$ and $(2\pi,0)$.

Expanding near these boundaries and matching to the interior gives
integer powers of $\mu^{-2}$ with at most one logarithm at each
order. The mixed term starts at $\mu^{-4}$ and has no logarithms.
The Bessel moments give $f_1\to-1$, hence
\begin{equation}
 \EEC_{2,2}(1;\lambda_Q)\sim-2+
 \sum_{n=1}^{\infty}\frac{A_n\log\lambda_Q+B_n}{\lambda_Q^n},
 \qquad \mu^2=8\lambda_Q.
 \label{eq:app-sc-result}
\end{equation}
Here $A_n,B_n$ are the coefficients of the full $N=K=2$ EEC
expanded in $\lambda_Q$, including the multiplicity of two.
The first coefficients are
\begin{equation}
 A_1=A_2=0,\qquad B_1=\frac1{12\pi^2},\qquad
 B_2=-\frac1{240\pi^2}-\frac3{200\pi^4}.
\end{equation}
The first logarithm occurs at order $\lambda_Q^{-3}$.

\subsection{All-order coefficients}

For the squared images $\widetilde t_{1,0}^{\,2}$ and $\widetilde t_{1,-1}^{\,2}$, set $\chi=R_0^2 = \alpha\beta$ which we call the squared image distance. Integration over $\varphi$ defines the mass-independent kernel
\begin{align}
 E_4(\chi)&=\operatorname{FP}
 \int_{\sqrt\chi}^{\pi+\chi/(4\pi)}
 \frac{\varphi^2}{\sqrt{\varphi^2-\chi}}
 \frac{[\cos\sqrt{\varphi^2-\chi}-\cos\varphi]^4
       (1+2\cos^2\varphi)}
      {\sin^6\varphi}\,d\varphi\notag\\
 &\sim\sum_{j\geq2}\chi^j
       (e_{4,j}+d_{4,j}\log\chi).
 \label{eq:app-sc-kernelE}
\end{align}
The remaining mass dependence is carried by the Bessel functions:
\begin{equation}
 f_1+1\sim\frac{2\mu^2}{\pi^2}\operatorname{FP}
 \int_0^{4\pi^2}\frac{d\chi}{\chi}\,
 K_1(\mu\sqrt\chi)^2E_4(\chi)
 +f_{1,\mathrm{cross}}(\mu).
\end{equation}
Here $f_{1,\mathrm{cross}}$ comes from
$2\widetilde t_{1,0}\widetilde t_{1,-1}$.
We see that the small-$\chi$ coefficients of $E_4$ and the mixed-image contribution determine $A_n,B_n$. Writing
\begin{equation}
 \widetilde t_{1,k}
 =2h\mu\frac{\varphi+2\pi k}{R_k}K_1(\mu R_k),
\end{equation}
we define the mixed-image coefficients by
\begin{equation}
 \frac{\pi^2}{2}f_{1,\mathrm{cross}}(\mu)
 =\frac18\operatorname{FP}\int_0^{2\pi}\!d\alpha
       \int_0^{2\pi}\!d\beta\,
       \mathcal W\widetilde t_{1,0}\widetilde t_{1,-1}
 \sim\sum_{n\geq2}\frac{\Xi_n}{\mu^{2n}},\qquad \Xi_1=0.
\end{equation}
With the finite part in~\eqref{eq:app-sc-finite-part},
$\Xi_2=6/(25\pi^2)$ and $\Xi_3=12/(5\pi^4)$.
The substitution $\chi=\rho^2/\mu^2$ reduces the mass integrals to
\begin{equation}
 \mathcal M_{n+1}
 =2\int_0^\infty\rho^{2n+1}K_1(\rho)^2\,d\rho
 =\frac{2^{2n}(n!)^2(n+1)!(n-1)!}{(2n+1)!},\qquad n\geq1.
\end{equation}
The logarithmic moments satisfy
\[
 2\int_0^\infty\rho^{2n+1}K_1(\rho)^2\log\rho^2\,d\rho
 =\mathcal M_{n+1}(2\log2+\Lambda_n),
\]
where
\begin{equation}
 \Lambda_n=-2\gamma_{\rm E}
 +2H_n+H_{n+1}+H_{n-1}-2H_{2n+1},
\end{equation}
and $H_n=\sum_{j=1}^n j^{-1}$, with $H_0=0$.
Using $\mu^2=8\lambda_Q$, we obtain
\begin{align}
 A_n&=-\frac{2^{2-3n}}{\pi^2}
       \mathcal M_{n+1}d_{4,n+1},\nonumber\\
 B_n&=\frac{2^{2-3n}}{\pi^2}\Bigl[
 \mathcal M_{n+1}\bigl(e_{4,n+1}+
 [\Lambda_n-\log2]d_{4,n+1}\bigr)+\Xi_n\Bigr].
 \label{eq:app-sc-AB}
\end{align}
These representations generate the formal expansion at arbitrary
order from the small-$\chi$ coefficients of the specified kernel.

\subsection{Scalar-detector endpoint}

The scalar integral $\mathcal F$ before applying $\mathbb D_\xi$
also has a compact endpoint representation. We retain
$\mu=2M_1=\sqrt{8\lambda_Q}$ and introduce triangle coordinates
$a,b$, distinct from the compact-contour coordinates $\alpha,\beta$.
Define the normalized triangle average
\begin{equation}
 \langle G\rangle_\triangle
 =\frac{2}{\pi^2}\int_0^\pi\!da
   \int_0^{\pi-a}\!db\,G(a,b),
 \label{eq:results-scalar-triangle-average}
\end{equation}
with propagator
\begin{equation}
 \widehat t_1(a,b;\lambda_Q)
 =\frac{2(\cos a+\cos b)}{\sin b}
 \int_0^\infty\!dp\,\cosh(ap)
 \frac{\sinh\!\left(b\sqrt{p^2+\mu^2}\right)}
      {\sinh\!\left(\pi\sqrt{p^2+\mu^2}\right)}.
 \label{eq:results-scalar-endpoint-propagator}
\end{equation}
Then
\begin{equation}
 \mathcal F_{2,2}(1;\lambda_Q)
 =\frac12\bigl\langle\widehat t_1^{\,2}-1\bigr\rangle_\triangle.
 \label{eq:results-scalar-endpoint-finite-coupling}
\end{equation}
At zero coupling $\widehat t_1(a,b;0)=1$.
In the open triangle $\widehat t_1\to0$ at strong coupling,
and $0\leq\widehat t_1\leq1$. Since
$\langle1\rangle_\triangle=1$, the scalar endpoint tends to $-1/2$. 

\endgroup
\bibliographystyle{JHEP}

\bibliography{refs}

\end{document}